\documentclass[twocolumn, switch]{article} 

\usepackage{preprint}

\usepackage{amsmath, amsthm, amssymb, amsfonts}

\usepackage[numbers,square]{natbib}
\usepackage[utf8]{inputenc}	
\usepackage[T1]{fontenc}	
\usepackage{xcolor}		
\usepackage[colorlinks = true,
            linkcolor = purple,
            urlcolor  = blue,
            citecolor = cyan,
            anchorcolor = black]{hyperref}	
\usepackage{booktabs} 		
\usepackage{nicefrac}		
\usepackage{microtype}		
\usepackage{lineno}		
\usepackage{float}			
\usepackage{xurl}
\usepackage{longtable}
\usepackage{stfloats} 

\usepackage{newfloat}
\DeclareFloatingEnvironment[name={Supplementary Figure}]{suppfigure}
\usepackage{sidecap}
\sidecaptionvpos{figure}{c}

\usepackage{graphicx} 
\usepackage{pgf-umlsd} 

\usepackage{tikz,xcolor,hyperref}

\usepackage{geometry} 
\usepackage{algorithm}
\usepackage{algpseudocode}
\usepackage{refcount}
\usepackage{caption}
\usepackage{mathtools} 
\usepackage{bm} 

\usepackage{array}
\newcolumntype{L}[1]{>{\raggedright\let\newline\\\arraybackslash\hspace{0pt}}m{#1}}
\newcolumntype{C}[1]{>{\centering\let\newline\\\arraybackslash\hspace{0pt}}m{#1}}
\newcolumntype{R}[1]{>{\raggedleft\let\newline\\\arraybackslash\hspace{0pt}}m{#1}}

\definecolor{lime}{HTML}{A6CE39}
\DeclareRobustCommand{\orcidicon}{
	\begin{tikzpicture}
	\draw[lime, fill=lime] (0,0)
	circle [radius=0.16]
	node[white] {{\fontfamily{qag}\selectfont \tiny ID}};
	\draw[white, fill=white] (-0.0625,0.095)
	circle [radius=0.007];
	\end{tikzpicture}
	\hspace{-2mm}
}
\foreach \x in {A, ..., Z}{\expandafter\xdef\csname orcid\x\endcsname{\noexpand\href{https://orcid.org/\csname orcidauthor\x\endcsname}
			{\noexpand\orcidicon}}
}

\title{A Regulatory Compliance Protocol for Asset Interoperability Between Traditional and Decentralized Finance in Tokenized Capital Markets}

\usepackage{authblk}

\author[1,2]{Jinwook Kim}
\author[1,2]{Oraclizer Research Group}
\author[1,2\thanks{\tt{jonghun@oraclizer.io}}]{Jonghun Hong}

\affil[1]{Oraclizer Labs, Delaware, USA}
\affil[2]{Oraclizer Labs Korea, Seoul, Korea}

\newcommand{\paperfirstversion}{March 26, 2024}
\date{July 3, 2026}
\renewcommand{\today}{July 3, 2026}

\begin{document}

\twocolumn[ 
  \begin{@twocolumnfalse} 

\maketitle
\setcounter{footnote}{0}

\begin{abstract}
There have been various attempts at token standards on numerous blockchain platforms today to fundamentally change the way assets are traded in the traditional capital markets, but there is a lack of research and resolution on regulatory issues that become the common foundation for interoperability and reusable standards. Our proposal, Regulatory Compliance Protocol (RCP), is based on the regulations and reports of 15 global financial institutions and standardizes recommendations and guidelines involving the overall asset tokenization of TradFi and DeFi into five regulatory groups: Traceability, Privacy, Enforceability, Finality and Tokenizability, compiling them into 31 items and presenting a benchmark for technology and standards as an underlying protocol. To review the legality and effectiveness of RCP, it was validated based on three tokenization and trading scenarios, and by benchmarking existing asset-tokenization standards (ERC-20, ERC-7943, ERC-1400, and ERC-3643) against RCP, it makes explicit which regulatory requirements each standard addresses at the token level and which remain inherently off-chain.
\end{abstract}
\vspace{0.35cm}

  \end{@twocolumnfalse} 
] 



\section{Introduction}
Interest in tokenized assets, from the tokenization of traditional finance (TradFi) assets to decentralized finance (DeFi) interoperability, has significantly increased, reflecting the broader trend towards digital innovation in the finance and web3 industries. The process of converting real assets into digital tokens on Distributed Ledger Technology (DLT) platforms offers numerous benefits, including enhanced liquidity and transparency, fractional ownership, and improved capital efficiency and accessibility. The surge in interest can be evidenced by the increase in academic articles, market reports, and the flow of investments into asset tokenization-specific projects and startups. Additionally, regulatory bodies and financial institutions have started to recognize the potential of asset tokenization in paving the way for a more efficient financial ecosystem. AN ASSESSMENT ON THE BENEFITS OF BOND TOKENIZATION\cite{1HKMA2023ASSESSMENT} The interest of financial regulators in tokenization technology underscores the central role asset tokenization is expected to play in the evolution of capital markets.

Compliance with global financial regulatory bodies is the cornerstone of asset tokenization within the capital markets sector, and it represents an indispensable fundamental requirement. AN ASSESSMENT ON THE BENEFITS OF BOND TOKENIZATION\cite{1HKMA2023ASSESSMENT} The essence of asset tokenization necessitates adherence to complex regulations regarding issuance, trading, and auditing to ensure the legality, security, and trustworthiness of tokenized assets. The regulatory frameworks of global financial regulatory bodies are designed to protect investors, maintain the integrity of the financial system and markets, and prevent financial crimes. Therefore, all tokenization schemes, from the TradFi industry to DeFi ecosystems interoperating with tokenized financial instruments from TradFi, must meticulously comply with existing legal standards. Failure to adhere to these regulations can not only compromise the legality of the tokenized assets but also expose the involved parties to legal risks and potential financial penalties. In conclusion, the path to comprehensive asset tokenization is inseparable from a thorough understanding and application of regulatory mandates, with compliance being not only essential for success in the evolving digital asset environment but also a prerequisite.

\begin{figure}
  \centering
  \includegraphics[scale=0.45]{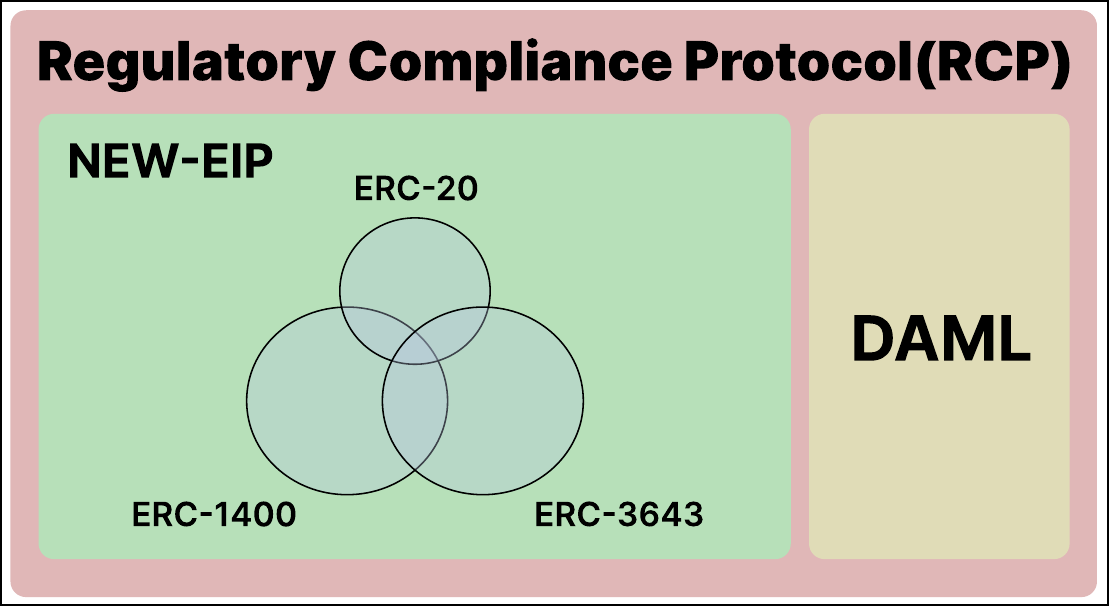}
  \caption{Regulatory Compliance Protocol}
  \label{fig:fig1}
\end{figure}

In the evolving blockchain technology landscape, protocols such as ERC-1400 Answering the Need for Standardization\cite{2POLYMATH2018Answering} and ERC-3643 Whitepaper ERC3643 The T-REX protocol\cite{3Tokeny2023ERC-3643}  have made significant progress in financial regulatory compliance related to asset tokenization in the DeFi ecosystem. These financial regulation-related proposals are designed to facilitate the issuance, control, and management of security tokens on Ethereum Virtual Machine (EVM\footnote{Ethereum Virtual Machine (EVM): A virtual execution environment that executes and processes smart contracts based on the consensus of the blockchain network}) compatible blockchains, including functions to address regulatory requirements for tokenized assets. However, ERC-1400 and 3643 face limitations in the completeness of financial regulation related to asset tokenization. The fundamental issue lies in the imperfect alignment between the regulatory and guideline provisions of various financial regulatory bodies that have been maturing and evolving over a long period in the global capital markets and the autonomous regulatory track of the DeFi ecosystem. This misalignment of regulatory tracks inherently hinders capital liquidity and asset interoperability between the TradFi industry and the DeFi ecosystem due to the uncertain legal risks associated with asset management. Therefore, a protocol that can integrate the regulatory tracks of both parties into a common framework is essential.

Our proposed protocol is a Regulatory Compliance Protocol (RCP) that ensures integrated compliance with financial regulations affecting asset interoperability between TradFi and DeFi. RCP serves as the underlying protocol for executable protocols in the tokenized capital markets, eliminating legal uncertainties in asset management and facilitating asset tokenization and capital liquidity. This includes standardizing the recommendations of various regulatory bodies into Traceability, Privacy, Enforceability\footnote{The characteristic of having binding force and being enforceable under defined conditions of laws, regulations, or policies}, Finality and Tokenizability\footnote{The ability to structure tokens according to the unique commodity guidelines and rules of various asset classes}.  It encompasses fundamental settings and rules for financial products, including identity verification and screening based on the Risk Based Approach (RBA)\footnote{The methodology of identifying, evaluating, and prioritizing potential risks in a specific activity or process, and allocating resources based on this to manage risks} principles of the TradFi industry, freezing of assets through regulatory audits, restrictions on access and transfer, and controls such as cancellation, modification, and setting limits on transactions. The RCP we have designed sets a new benchmark for the seamless interoperability of tokenized assets within the digital ecosystem, fully complying with complex financial transaction regulations.

RCP serves as a regulatory bridge for the tokenized capital markets, facilitating true financial integration between the existing capital markets and the rapidly growing web3 ecosystem. Unlike the existing ERC-1400 and 3643 protocols, RCP will delve deeper into existing financial regulations and complexities to maximize the potential of asset tokenization through protocols, laying the groundwork for research aimed at leveraging the potential of asset tokenization. This research, by resolving regulatory uncertainties in the interoperability between real assets and digital native assets\footnote{Assets that are issued in digital form and exist solely in the digital realm}, provides a technical basis for securely and atomically completing transactions between the two, enabling stakeholders to enjoy enhanced trust, liquidity, global market access, and the benefits of fractional ownership within the RCP framework. For example, by tokenizing real estate in the TradFi industry and allowing global investors through DeFi to partially own and trade shares of that real estate in their digital wallets, RCP's compliance with finality can protect investors while broadening access to investment opportunities and stimulating economic activity. Therefore, RCP not only represents a benchmark for the technical foundation of asset interoperability in the tokenized capital markets but also heralds a new era of financial innovation that brings the diverse assets of our world closer together through the expanded use of asset tokenization.

\section{Theoretical Background}
\label{sec:headings}

In the blockchain protocol domain, ERC-1400 and 3643 have shown significant progress in addressing compliance issues. However, their approaches are diverse and, crucially, they have not fully met the comprehensive standards recommended by regulatory bodies. The essence of these protocols lies in the attempt to standardize the tokenization process through integrating specific regulatory compliance mechanisms tailored to the digital native assets realm. Despite these efforts, discrepancies arise when juxtaposed with the broad regulatory frameworks established by financial supervisory bodies. The primary issue with existing protocols like ERC-1400, 3643 is that their scope and depth are insufficient to encapsulate the full spectrum of regulatory guidelines related to asset tokenization. Regulatory bodies support a more holistic and comprehensive approach that not only addresses the digital dimensions of assets but also intertwines the legal and operational nuances of the existing financial system. To bridge this gap and unify the direction of regulatory compliance while ensuring a volume of compliance robust enough to meet the stringent requirements set by regulatory bodies, a regulatory compliance protocol is needed. The development of such standards is paramount in ensuring regulatory compliance for asset tokenization and trading, creating a safe and efficient environment, and represents a critical step in integrating DLT with the traditional financial ecosystem.

\subsection{ERC-1400 Protocol}
The ERC-1400 protocol has served as an innovative beacon in the blockchain domain, heralding a new era of standardization for security tokens. Designed as an EVM-based smart contract interface to meet the complex financial transactions and regulatory compliance requirements, ERC-1400 provides a standardized framework for security tokens. It is compatible with existing token standards such as ERC-20 and ERC-777, while supporting divisible security tokens, transaction restrictions, and document management, thus offering a foundational environment for the diversification of financial products and regulatory compliance.

While the ERC-1400 protocol has established itself as a robust foundation for security token transactions in the blockchain domain, simulating the tokenization process for various financial products reveals significant limitations.The core issue lies in whether the protocol meets or fails to meet the stringent recommendations and product guidelines set by regulatory bodies managing tokenized assets. Despite its innovative approach to digital securities, ERC-1400 faces structural limitations in flexible management of customer identity, suspicious transaction monitoring, transaction cancellation or modification, contract-level suspension and disposal of financial products, blacklist management, and the inability to set token expiration. These requirements are crucial for ensuring the legality, security, and transparency of tokenized assets, protecting stakeholders from fraud, and ensuring global financial law compliance.

\subsection{ERC-3643 Protocol}
The ERC-3643 protocol is a securities token standard of the Ethereum Request for Comments (ERC) designed with the tokenized capital markets in mind, aiming for compliance with financial regulations that tokenized financial products may be subject to. Building on the existing ERC-20 token standard, ERC-3643 incorporates various regulatory compliance features essential for the nature of financial products in a regulatory-intensive financial environment. These features enable asset tokenization through customer identity verification (KYC), asset freezing and retrieval, transaction restrictions and cancellations, token burning, and supply control, catering to the necessities of regulated financial activities.

The ERC-3643 protocol serves as a robust framework for blockchain-based transactions, particularly in the context of security tokens, but it has limitations in fully reflecting the complex requirements and regulatory environments of various financial regulatory bodies. While ERC-3643 focuses on the fundamental aspects of issuing and managing security tokens, it falls short in some enforcement and lifecycle needs of regulatory bodies, such as attaching and verifying legal documents, setting expiration dates for tokens, suspicious transaction monitoring, and managing differentiated asset classes. These omissions are crucial for complying with financial regulations and meeting the demands of diverse financial product guidelines in the capital markets. The discrepancy between the functionalities of ERC-1400, ERC-3643, and the recommendations of regulatory bodies underscores the need for more customized regulatory compliance protocols. A protocol that satisfies better financial regulations not only bridges the gap between the digital efficiencies offered by DLT and the complex legal environment of real assets but is also essential for protecting and stabilizing the market.

\subsection{Financial Regulator}

\begin{table*}[t]
  \centering
  {\tiny
  \begin{tabular}{R{0.4cm}  C{0.7cm}  C{0.7cm}  C{0.7cm}  C{0.7cm}  C{0.7cm}  C{0.7cm}  C{0.7cm}  C{0.7cm}  C{0.7cm}  C{0.7cm}  C{0.7cm}  C{0.7cm}  C{0.7cm}  C{0.7cm}  C{0.7cm}}
    \toprule
    RCP & WB & ISDA & IOSCO & IMF & FSB & FATF & BIS & SFC & HKMA & EU & ESMA & FCA & MAS & FINMA & FINRA\\
    \midrule
    (1) & \checkmark & & \checkmark & \checkmark & \checkmark & \checkmark & \checkmark & \checkmark & \checkmark & \checkmark & & & & \checkmark & \\
    \hline
    (2) & & \checkmark & \checkmark & \checkmark & \checkmark & \checkmark & \checkmark & \checkmark & \checkmark & \checkmark & & & \checkmark & \checkmark & \checkmark \\
    \hline
    (3) & & & & & & \checkmark & & & \checkmark & \checkmark & & & & & \checkmark \\
    \hline
    (4) & & & & & & & & & \checkmark & & \checkmark & & & & \\
    \hline
    (5) & & \checkmark & \checkmark & & & & & \checkmark & & & & \checkmark & & & \\
    \hline
    (6) & & & \checkmark & \checkmark & \checkmark & & & \checkmark & & \checkmark & \checkmark & \checkmark & & \checkmark & \checkmark \\
    \hline
    (7) & & & & & & \checkmark & & & \checkmark & \checkmark & & & \checkmark & \checkmark & \\
    \hline
    (8) & & & & & & \checkmark & & & & & & & & & \\
    \hline
    (9) & & & & & & \checkmark & & & & & & & & & \\
    \hline
    (10) & & & \checkmark & \checkmark & \checkmark & \checkmark & & \checkmark & & \checkmark & & & & & \checkmark \\
    \hline
    (11) & & \checkmark & & & & \checkmark & & \checkmark & & & & & & \checkmark & \\
    \hline
    (12) & & \checkmark & \checkmark & & & & \checkmark & & & & & & & & \checkmark \\
    \hline
    (13) & & \checkmark & \checkmark & & & \checkmark & & \checkmark & & \checkmark & & & & & \checkmark \\
    \hline
    (14) & & & & & & & & & & \checkmark & & & & & \\
    \hline
    (15) & & & & & & \checkmark & & \checkmark & & & & & & & \checkmark \\
    \hline
    (16) & & & \checkmark & & & & \checkmark & & & & & & & & \checkmark  \\
    \hline
    (17) & \checkmark & \checkmark & \checkmark & & & \checkmark & & \checkmark & \checkmark & \checkmark & & & \checkmark & & \checkmark    \\
    \hline
    (18) & \checkmark & & \checkmark & & & \checkmark & & & \checkmark & & \checkmark & & & &    \\
    \hline
    (19) & & & & & & & & & & \checkmark & & & & &    \\
    \hline
    (20) & & & \checkmark & & & & \checkmark & & \checkmark & & \checkmark & & & &    \\
    \hline
    (21) & & \checkmark & \checkmark & & & \checkmark & \checkmark & \checkmark & \checkmark & \checkmark & & \checkmark & & & \checkmark    \\
    \hline
    (22) & & & & & & & & & \checkmark & \checkmark & & & & & \checkmark    \\
    \hline
    (23) & & & & & & & & & & & & & & & \checkmark    \\
    \hline
    (24) & & & & & & & & & & & \checkmark & & & & \checkmark    \\
    \hline
    (25) & & & \checkmark & & & & & & & & & & & &    \\
    \hline
    (26) & & & & & & \checkmark & & & & & \checkmark & & & &    \\
    \hline
    (27) & & & & & & & & & & & & & & &    \\
    \hline
    (28) & & & \checkmark & & & & \checkmark & & & & & & & & \checkmark   \\
    \hline
    (29) & & & & & & & & & & & & & & &    \\
    \hline
    (30) & & & \checkmark & & & & & & & & & & & &    \\
    \hline
    (31) & & & & & & & & & & & & & & &    \\

    \bottomrule
  \end{tabular}
  }
  \label{tab:table}
  \caption{Recommendation and Guidance of Regulatory Authorities}
  \tiny
  (1) Customer Identity Verification (2) High-Risk/Suspicious Transaction Monitoring (3) Detection of Changes to Customer Identity Information (4) Contract Version Tracking (5) Exploration of Transaction History by Asset Type (6) External Audit (7) Setting Role-Based Permissions (8) Asset Freeze (9) Asset Recovery (10) Trading Restrictions (11) Transaction Limit (12) Cancellation or Modification of Transactions (13) Pausing of Trading (14) Suspension or Disposal of Smart Contract (kill switch) (15) Blacklist Management (16) Forced Liquidation (17) Privacy of Personal Information (18) Privacy of Financial Transactions (19) Code Security (20) Immutability of the Ledger (21) Finality of Transactions and Payments (22) Attaching Legal Documents (23) Token Expired Time (24) Token Transfer Restrictions (25) Issuance of Tokenized Cash (26)  Issuance of Tokenized Securities  (27) Controlling Transactions Involving Splitting Below Decimal Units (28) Token Burning (29) Gasless Support (30) Asset Class Management (31) Token Supply Control
\end{table*}

In the financial industry, asset tokenization refers to the process of converting assets from traditional capital markets into digital form through tokenization technologies such as DLT, hence the market where these tokenized digital assets are traded is also considered part of the capital markets. Financial regulatory bodies aim to ensure that tokenized assets are treated in a similar manner to traditional financial products, pursuing goals such as protecting investors, maintaining market transparency and integrity, and preventing financial crimes, thereby preserving and advancing market order. Regulatory bodies are categorized into international financial regulatory and standard-setting bodies, financial market and product regulatory agencies, and national regulatory agencies.

Leading international financial regulatory and standard-setting bodies such as the IMF, BIS, FATF, and FSB play a pivotal role in shaping the global financial environment. The International Monetary Fund (IMF) leads in promoting global macroeconomic and financial stability by providing policy advice and capacity development support aimed at fostering a strong and sustainable global economy. The Bank for International Settlements (BIS) is crucial in implementing monetary policies and regulating and supervising banks to ensure the stability of the financial system. The Financial Action Task Force (FATF) leads in establishing international standards to prevent money laundering, terrorist financing, and proliferation financing, thereby strengthening global security and integrity. The Financial Stability Board (FSB) integrates efforts of national financial authorities and international standard-setting bodies to develop and promote consistent application of regulatory, supervisory, and financial sector policies across jurisdictions. Collectively, these organizations embody the critical characteristics of international financial regulatory and standard-setting bodies by focusing on stability, integrity, collaboration, and resilience. They strive to mitigate risks, enhance transparency, and promote international cooperation and harmony in financial regulation, ensuring a safer and more stable global financial environment.

Regulatory bodies in the financial markets and products sector, among them ISDA and IOSCO, play a central role in setting standards and developing best practices, mitigating risks, protecting investors, and enhancing market fairness in the financial markets and products domain. These entities facilitate the creation of guidelines, frameworks, and widely accepted standards designed to improve market efficiency and effectiveness, thereby setting benchmarks for global securities and financial transactions. Prioritizing the identification and reduction of risks inherent in financial products, as seen in International Swaps and Derivatives Association(ISDA)'s efforts to streamline derivative transactions, their mission's cornerstone is to protect investors' interests and ensure fair market operations. They also emphasize the importance of cross-border cooperation, striving for regulatory consistency across jurisdictions to facilitate smooth global asset interoperability. Through efforts in capital market integration, systematic threat reduction, and provision of educational resources and guidelines, these regulatory bodies play a vital role in fostering a stable, transparent, and efficient financial ecosystem that supports economic growth and development worldwide.

National institutions proactive in digital asset regulation, among them the SFC, HKMA, ESMA, FCA, MAS, FINMA, and FINRA, play a crucial role in shaping the regulatory environment for cryptocurrencies and other digital financial products. The Securities and Futures Commission( SFC) in Hong Kong conducts comprehensive supervision of the securities and futures markets based on extensive investigation and enforcement capabilities. The Hong Kong Monetary Authority (HKMA), acting as Hong Kong's central bank, supervises financial institutions and ensures financial stability. European Securities and Markets Authority (ESMA) strives to stabilize and rationalize EU financial markets while enhancing transparency and protecting investors. The Financial Conduct Authority (FCA) in the UK focuses on protecting consumer interests, enhancing market efficiency, and ensuring financial stability. Monetary Authority of Singapore (MAS), serving both as Singapore's central bank and financial regulatory authority, plays a pivotal role in maintaining the robustness of Singapore's financial system. Swiss Financial Market Supervisory Authority(FINMA) ensures market transparency and supervises market participants in Switzerland. Financial industry regulatory authority (FINRA) in the US is dedicated to protecting investors, ensuring market integrity, and fostering fair and efficient markets.

These institutions are increasingly undertaking innovative tasks to integrate digital assets into the existing financial framework, emphasizing the importance of innovation, consumer protection, market integrity, and Anti-Money Laundering (AML) compliance in the rapidly evolving digital financial landscape. Their efforts include developing specific guidelines for digital asset transactions, supervising digital asset service providers, and implementing technologies to monitor and regulate digital financial markets. By actively adjusting regulatory approaches to accommodate the unique aspects of digital assets, these national agencies aim to protect investors and ensure market fairness while creating an environment that supports the growth of digital finance and its integration into the broader financial ecosystem.

\section{Recommendations and Guidance from Regulatory Agencies}

Requirements based on the recommendations of financial regulatory bodies are more comprehensive and stringent than regulations in other asset areas that can be tokenized, surpassing the inherent functionalities of blockchain. In the interoperability between tokenized assets divided into different domains such as Traditional Finance (TradFi) and Decentralized Finance (DeFi), compliance with financial regulations becomes the starting point for asset liquidity. However, due to the varying content of regulations by institutions and the lack of attempts and research to integrate these complex regulations, there has been difficulty in fully complying. Therefore, we aimed to create protocols that comply with all the regulations foundational for asset interoperability in the tokenized capital markets by consolidating and standardizing the content of regulations and guidelines by institution and nature. In this chapter, we detail the elements of regulations included in the RCP, divided by nature and recommendation. The specific provisions of each institution regarding the regulations are attached in the Table 4 of the appendix chapter.

\subsection{Traceability}

Traceability, extracted through a comprehensive analysis of recommendations and guidelines from various global standard-setting and regulatory bodies, is a critical attribute for maintaining a robust token infrastructure. This includes systematic recommendations for identifying, tracking, and verifying the history, distribution, and location of assets within the network. This attribute facilitates compliance with Know Your Customer (KYC) regulations, ensuring financial safety through Anti-Money Laundering (AML) and Counter-Financing of Terrorism (CFT). Various regulatory bodies such as the World Bank, FINMA, HKMA, and FATF emphasize these recommendations. Traceability requires mechanisms for customer due diligence, monitoring suspicious transactions, detecting changes in customer identity information, and transparent auditing and reporting of transaction activities. Through these mechanisms, financial institutions and providers of tokenization services can mitigate risks, maintain financial integrity, and ensure compliance with global standards. Therefore, traceability not only forms the basis for legal and regulatory compliance but also enhances the security and reliability of the tokenized asset ecosystem, fostering trust between participants and regulatory bodies.

\paragraph{Customer Identity Verification}
Regarding customer identity verification, FATF emphasized the necessity of performing Customer Due Diligence (CDD) in The FATF Recommendations\cite{15FATF2023Recommendations}, stating that "financial institutions should be required to identify and verify the identity of the customer and understand the nature of its business and its ownership and control structure." Similarly, FINMA described in the Verordnung der Eidgenössischen Finanzmarktaufsicht über die Bekämpfung von Geldwäscherei und Terrorismusfinanzierung im Finanzsektor\cite{17FINMA2023Verordnung}, "When establishing a business relationship with a natural person or a sole proprietor, the financial intermediary identifies the contracting party by examining an identification document provided by the contracting party." Additionally, HKMA highlighted the importance of managing Digital ID (D-ID) to simplify the KYC process in its Whitepaper 2.0 on Distributed Ledger Technology\cite{20HKMA2017Ledger}.
These provisions from various regulatory bodies underline the absolute necessity of customer identity verification. They emphasize that a strong mechanism for verifying customer identity is essential to maintain the safety of financial transactions, prevent money laundering, and stop the financing of terrorism. This function serves as the foundation for complying with international regulatory standards and creating a reliable and safe financial environment.

\paragraph{High-Risk/Suspicious Transaction Monitoring}
FATF stated in The FATF Recommendations\cite{15FATF2023Recommendations} that "All suspicious transactions including attempted transactions should be reported regardless of the amount of the transaction," emphasizing the importance of monitoring and reporting suspicious activities. This recommendation plays a pivotal role in detecting and preventing illegal financial flows. Similarly, FATF underscored the importance of continuous vigilance by stating, "Ongoing monitoring on a risk basis means scrutinizing transactions to determine whether those transactions are consistent with the VASP’s (or other obliged entity’s) information about the customer and the nature and purpose of the business relationship." Additionally, MAS advocated for proactive measures in its Technology Risk Management Guidelines\cite{28MAS2021Technology} by stating, "The FI should implement real-time fraud monitoring systems to identify and block suspicious or fraudulent online transactions."

These direct provisions from authoritative sources like FATF and MAS reveal the importance of monitoring high-risk and suspicious transactions within the financial industry. They clearly demonstrate that stringent and continuous monitoring is essential for identifying, reporting, and taking action on suspicious activities, which is crucial for the safety of the financial system and for preventing money laundering and terrorist financing. Ultimately, these recommendations emphasize the need to adopt sophisticated monitoring mechanisms to implement a commitment to a safe, transparent, and compliant financial ecosystem and protect against financial crimes.

\paragraph{Detection of Changes to Customer Identity Information}
FATF emphasizes the importance of keeping customer information current in The FATF Recommendations\cite{15FATF2023Recommendations}, specifically stating that financial institutions "should undertake reviews of existing records to keep documents, data, or information collected under the CDD process up-to-date and relevant." This is also reflected in FINRA's FINRA Rules\cite{18FINRA2018Rules}, which require daily updates of customer information and ensure that all changes to the customer profile are accurately and promptly recorded to maintain the safety of financial transactions.

These provisions underscore the significance of detecting changes in customer identity information within the financial ecosystem. They highlight a collective understanding among regulatory bodies about the need for ongoing vigilance in monitoring and updating customer-related data. Such practices are crucial not only for preventing financial crimes such as money laundering and the financing of terrorism but also for ensuring the reliability of the broader financial system. By mandating that financial institutions actively manage and update customer information, the goal is to foster an environment of significant trust, transparency, and security in the global financial market.

\paragraph{Contract Version Tracking}
ESMA emphasized the importance of all participants maintaining an identical, up-to-date ledger in their guidelines on Advice, Initial Coin Offerings and Crypto-Assets\cite{8ESMA2019Initial}, stating, "Each party who participates in the validation process has an identical, up-to-date copy of the chain or public ledger, which is a record of all the transactions." This principle is crucial for ensuring the integrity and verifiability of transactions on DLT. In the context of regulatory compliance, ESMA's Report on the DLT Pilot Regime, On the Call for Evidence on the DLT Pilot Regime and compensatory measures on supervisory data\cite{9ESMA2022Regime} clearly states, “DLT infrastructures that do not request the reporting exemption should have systems in place to ensure that the right sequencing is respected." This directly highlights the importance of tracking contract versions to maintain accurate and chronological transaction records.

ESMA's direct provisions have emphasized the importance of contract version tracking in the digital finance ecosystem. By maintaining accurate, up-to-date, and correctly sequenced records of transactions and contract versions, the integrity of financial transactions can be supported. These recommendations are particularly essential for the safe and efficient operation of DLT, enabling clear audit trails and ensuring the reliability of transaction histories, thereby fostering trust between market participants and regulatory bodies.

\paragraph{Exploration of Transaction History by Asset Type}
FCA mentioned the necessity of clarity in transaction history in their Finalised non-handbook guidance on Crypto Asset Financial Promotions\cite{14FCA2023Finalised} specifically stating, "firms should clearly and prominently disclose ‘who’ owns the legal and beneficial rights to the crypto asset as part of the financial promotion." This guidance emphasizes the importance of asset ownership and transaction history transparency. SFC, in its Guidelines for Virtual Asset Trading Platform Operators\cite{29SFC2023Guidelines}, mandated, "A Platform Operator should provide to each client timely and meaningful information about the transactions conducted with the client or on the client’s behalf," highlighting the need for detailed transaction history per asset type for consumer protection and transparency. Similarly, ISDA emphasized in LEGAL GUIDELINES FOR SMART DERIVATIVES CONTRACTS: THE ISDA MASTER AGREEMENT\cite{27ISDA2019MASTER} the importance of identifying payment streams, stating, "An important task in developing technology solutions will be to identify each of these potential payment streams...and how these payment streams might be affected by the provisions of the ISDA Master Agreement." This highlights the necessity of accurately differentiating and tracking transactions and payments related to various asset classes.

These direct provisions from FCA, SFC, and ISDA underscore the significance of the ability to navigate transaction histories by asset type in the financial industry. This functionality is crucial for ensuring transparency, facilitating regulatory compliance, and providing investors and stakeholders with the information needed to understand asset movements and ownership positions. Such capabilities not only increase trust in the capital markets but also ensure the integrity of transactions.

\paragraph{External Audit}
FINRA emphasizes the necessity of external audits in FINRA Rules\cite{18FINRA2018Rules}, stating that firms "submit an Auditor's Report to the SEC staff, which is not deemed unacceptable by the SEC staff." This requirement highlights the importance of external audits in verifying the integrity of financial practices and compliance. The EU mentions in REGULATION (EU) 2022/858\cite{12EU2022(EU)2022/858}, "The competent authority for a DLT market infrastructure should be allowed to require an audit to ensure that the overall IT and cyber arrangements of the DLT market infrastructure are fit for purpose," emphasizing the importance of audits in assessing the service purpose and technical reliability of DLT systems. Furthermore, the IMF and FSB in Synthesis Paper: Policies for Crypto-Assets\cite{19IMF-FSB2023Synthesis} present the importance of regulatory compliance through standard implementation and imply the role of external audits in such standard enforcement, stating, "Even when the standards are effectively implemented, regulators will need to actively monitor market developments and emerging vulnerabilities, as well as assess illicit finance risks."

These provisions particularly demand the necessity of external audits in the digital asset ecosystem, related to regulatory compliance, operational integrity, and technological soundness. External audits serve as an indispensable tool in ensuring transparency, accountability, and reliability between financial institutions and technology providers, thereby supporting the stability and security of the global financial system.

\subsection{Privacy}

In the context of tokenization infrastructure technology, privacy serves as one of the fundamental information security principles of traditional finance aimed at protecting sensitive information from being exposed on fully public ledgers. This principle is underscored by various regulatory frameworks and guidelines that collectively advocate for the secrecy of financial transactions, protection of source code, and privacy practices. Achieving this requires complex measures that are challenging to implement on blockchain, such as strict access control settings based on roles and authority, encryption of contract code, and anonymization of sensitive personal information. Moreover, regulatory bodies recommend stringent compliance with privacy laws, including the "right to be forgotten," to address issues arising from the inherent transparency of blockchain that could lead to unintentional disclosure of participant identities. This holistic approach to privacy plays a crucial role in building trust and maintaining the information security of tokenization infrastructures, thereby enabling the sustainable development and widespread adoption of tokenized capital markets.

\paragraph{Privacy of Personal Information}
The EU emphasizes data protection measures for natural persons in the GDPR\footnote{GDPR(General Data Protection Regulation) : A law enacted by the European Union to strengthen the protection of personal data and privacy rights} (General Data Protection Regulation), (EU) 2016/679\cite{33EU2016GDPR}, recommending the principle "The principles of data protection should apply to any information concerning an identified or identifiable natural person." It also mentions the importance of the 'right to be forgotten' in "Article 17 Right to erasure (‘right to be forgotten’)" stating, "The data subject shall have the right to obtain from the controller the erasure of personal data concerning them." This strong adherence to data privacy principles underlines the need for meeting obligations of personal information and data protection globally. ISDA reinforces this principle in ISDA Legal Guidelines for Smart Derivatives Contracts: Foreign Exchange Derivatives\cite{26ISDA2020Exchange}, advising, "Only information that is permitted to be disclosed to each participant in the system (e.g., CCPs, regulators, brokers, parties) should be made available to them even where data is collected centrally." Additionally, the HKMA highlights the delicate balance between transparency in digital transactions and privacy in its Whitepaper 2.0 on Distributed Ledger Technology\cite{20HKMA2017Ledger}, recommending, "In addition to addressing the confidentiality of protected information stored in the DLT, it is important to consider the confidentiality of metadata stored in DLT."

These provisions collectively affirm the obligation of compliance with personal information and data privacy within the financial sector's regulatory framework. By advocating strict data segregation measures for confidential information, these guidelines spotlight the necessity of privacy in maintaining the financial system's integrity, protecting personal information, and adhering to global data protection standards. The concentrated regulatory focus by various authoritative bodies on privacy underscores its fundamental importance as a key element of safe, trustworthy, and compliant financial operations in the digital age.

\paragraph{Privacy of Financial Transactions(Data)}
The FATF states in The FATF Recommendations\cite{15FATF2023Recommendations} that "competent authorities should maintain appropriate confidentiality for any request for cooperation and the information exchanged" to protect the integrity of investigations and maintain privacy and data protection standards. This principle is also emphasized in the ESMA’s Advice, Initial Coin Offerings and Crypto-Assets\cite{8ESMA2019Initial}, mentioning "to guarantee the security and authentication of the means of transfer of information." Furthermore, the International Organization of Securities Commissions(IOSCO)  in Policy Recommendations for Crypto and Digital Asset Markets\cite{22IOSCO2023Policy} stated that CASPs (Cryptocurrency Asset Service Providers) must "put in place systems, policies, and procedures around the management of material non-public information."

These significant provisions from key financial regulatory bodies underscore the absolute necessity of privacy in financial transactions. By mandating stringent security measures, authentication protocols, and data protection policies, they highlight the fundamental role of privacy in protecting sensitive financial information, preventing misuse, and ensuring the integrity and reliability of the financial markets. The collective emphasis on privacy across these provisions reinforces the indispensability of a safe, transparent, and efficient financial ecosystem, thereby underscoring the importance of tokenization infrastructure in enhancing its integrity.

\paragraph{Code Security}
The EU emphasizes the importance of strong IT and cybersecurity measures for DLT infrastructures in REGULATION (EU) 2022/858\cite{12EU2022(EU)2022/858}, stating "DLT market infrastructures should have specific and robust IT and cyber arrangements related to the use of distributed ledger technology." Such arrangements must be "proportionate to the nature, scale, and complexity of the business plan of the operator of the DLT market infrastructure" and ensure "integrity, security, confidentiality, availability, and accessibility of data stored on the distributed ledger." This underscores the need to protect the confidentiality and security of contract codes and related data within DLT systems.

By mandating comprehensive confidentiality measures, including source code confidentiality within DLT systems, the EU sets a high standard for the protection of DLT infrastructures. This regulatory focus on source code security is increasingly important in building trust in decentralized ledger systems like DLT

\subsection{Enforceability}

Enforcement refers to the implementation of compliance and control measures by financial regulatory bodies to protect and regulate access, transactions, and activities related to not only traditional financial service providers but also digital asset and virtual asset service providers. This includes a wide range of regulatory mechanisms such as access control measures, asset freezing guidelines, transaction restrictions, transaction limits, and protocols for canceling or modifying transactions. Enforcement is not merely about restrictions and controls; it plays a pivotal role in maintaining safe, transparent, and compliant capital markets.

\paragraph{Setting Role-Based Permissions}
MAS emphasizes the principle of "least privilege" in its Technology Risk Management Guidelines\cite{28MAS2021Technology}, stating "Access rights and system privileges should be granted according to the roles and responsibilities of the staff, contractors, and service providers." Similarly, the HKMA highlighted in its Whitepaper 2.0 on Distributed Ledger Technology\cite{20HKMA2017Ledger}, "The system needs to allow for distinct levels of permission. It must allow users to specify the level of confidentiality for each transaction."

The inclusion of such provisions by the MAS and HKMA demonstrates the importance of role-based permissions as the foundation for the governance of information and financial systems. By stipulating that access and regulatory permissions strictly align with an institution's roles and responsibilities, these measures not only protect market supervision authority but also minimize operational risks. Collectively, these measures are absolutely necessary for establishing a secure financial transaction order within the digital and financial ecosystem, serving as a function to maintain the guidelines and rules for financial products through the implementation of role-based permissions.

\paragraph{Asset Freeze}
FATF recommends in The FATF Recommendations\cite{15FATF2023Recommendations} that "Countries should ensure that, in the context of processing wire transfers, financial institutions take freezing action and should prohibit conducting transactions with designated persons and entities as per the obligations set out in the relevant United Nations Security Council resolutions." Furthermore, in the interpretation note to Recommendation 6, it is specified that "Countries should also freeze without delay the funds or other assets—including VAs—of designated persons or entities and ensure that no funds or other assets—including VAs—are made available to or for the benefit of designated persons or entities."

These provisions by FATF underscore the absolute necessity of asset freezing mechanisms within the financial and digital asset sectors. By mandating the immediate freezing of assets related to designated individuals and entities, these measures serve as a strong deterrent against the financing of terrorism and money laundering. As FATF has outlined, the ability to swiftly implement financial sanctions is crucial for establishing the order of transactions in the global financial system and prevents the financial network from being misused for malicious activities. The explicit requirement for immediate action on asset freezing and prohibiting transactions with designated entities highlights the significant role of regulatory bodies in maintaining financial stability and protecting against threats to national and international security.

\paragraph{Asset Recovery}
FATF emphasizes the importance of asset recovery in The FATF Recommendations\cite{15FATF2023Recommendations} through an interpretive note advocating comprehensive measures for the confiscation of criminal property. One of the key provisions states, "Countries need a comprehensive range of measures, including legislative measures, available to confiscate criminal property and property of corresponding value." Furthermore, FATF emphasizes international cooperation in asset recovery, arguing that "Countries should take part in and actively support multilateral networks to better facilitate rapid and constructive international cooperation in asset recovery." 

These provisions highlight the crucial role of asset recovery within the broader context of preventing money laundering, war crimes, and the financing of terrorism. The emphasis on a comprehensive legislative framework for the confiscation of criminal assets, along with the encouragement of international cooperation, underscores the necessity of asset recovery mechanisms to disrupt the financial networks supporting criminal activities. Asset recovery is essential not only for depriving criminals of their illicit gains and deterring criminal activities but also for restoring these assets to their rightful owners or the state, thereby mitigating the economic impact of crime. FATF's focus on asset recovery enhances its importance in maintaining the integrity of the financial system and ensuring that crime does not pay, thereby upholding justice.

\paragraph{Trading Restrictions}
FINRA clearly stated the necessity for trading restrictions to maintain market integrity in FINRA Rules\cite{18FINRA2018Rules}, indicating "FINRA may impose from time to time such restrictions on option transactions or the exercise of option contracts in one or more series of options of any class which it determines are necessary in the interest of maintaining a fair and orderly market." Similarly, the EU's MiFIR (Markets in Financial Instruments and Amending Regulation), (EU) No 600/2014\cite{35EU2014MiFIR}, permits venues to constrain a client's transaction exposure by stating, "In order to limit the risk of exposure to multiple transactions from the same client, systematic internalisers shall be allowed to limit in a non-discriminatory way the number of transactions from the same client."

These provisions by FINRA and the EU explain the essential role of trading restrictions in capital markets. By granting regulatory authorities the power to impose trading restrictions, these measures are designed to prevent market manipulation, protect investors, and ensure a level playing field for all market participants. Particularly in the context of systematic internalisers under MiFIR, the ability to limit a single client's transaction exposure serves to contain concentration risk and preserve orderly quoting. The emphasis on maintaining a fair and orderly market underscores the importance of trading restrictions not only for the stability of the capital markets but also for the protection of investors and the integrity of financial transactions.

\paragraph{Transaction Limit}
FATF has set a clear threshold for transaction limits to prevent money laundering and terrorist financing in The FATF Recommendations\cite{15FATF2023Recommendations}, stating "The designated threshold for occasional transactions under Recommendation 10 is USD/EUR 15,000." In the same context, FATF specifies a lower threshold for cross-border wire transfers, emphasizing "Countries may adopt a de minimis threshold for cross-border wire transfers (no higher than USD/EUR 1,000)." This is complemented by the SFC's Guidelines for Virtual Asset Trading Platform Operators\cite{29SFC2023Guidelines}, which recommend, "Except for institutional and qualified corporate professional investors, a Platform Operator should set a limit for each client to ensure that the client’s exposure to virtual assets is reasonable with reference to the client’s financial situation."

The guidelines from FATF and SFC demonstrate the importance of transaction limits as regulatory tools within the financial ecosystem, especially in relation to preventing money laundering and terrorist financing. Setting thresholds for occasional transactions and cross-border wire transfers mitigates the risk of large-scale illicit financial flows and subjects transactions exceeding specific amounts to enhanced scrutiny. Similarly, SFC's guidelines on limiting client exposure to virtual assets aim to prevent excessive risk-taking, thereby protecting investors and maintaining market stability. These measures highlight the necessity of transaction limits as means to enhance regulatory compliance, protect financial stability, and safeguard the integrity of the global financial system.

\paragraph{Cancellation or Modification of Transactions}
FINRA highlights the procedural aspects of trade modifications or cancellations in FINRA Rules\cite{18FINRA2018Rules} by stating, "Members shall append the applicable trade report modifiers or indicators as specified by FINRA to all transaction reports." This is further detailed in ISDA's LEGAL GUIDELINES FOR SMART DERIVATIVES CONTRACTS: THE ISDA MASTER AGREEMENT\cite{27ISDA2019MASTER}, which allows for the termination of transactions under specific conditions: "The ISDA Master Agreement allows either party (or in certain scenarios both parties) to terminate transactions entered into under the ISDA Master Agreement upon the occurrence of an event of default or termination event."

These provisions from FINRA and ISDA underline the fundamental necessity of control mechanisms within capital markets that allow regulatory bodies to permit the cancellation or modification of transactions. By ensuring through regulatory bodies the ability to demand modifications of transactions in predefined circumstances and to terminate contracts, the aim is to maintain a high level of flexibility and responsiveness in the financial asset transaction process. This flexibility is crucial for resolving errors, managing risk, and responding to unforeseen events, thereby enhancing the resilience and integrity of the financial markets. The ability to adjust or discontinue transactions based on new information or changes in circumstances is essential to protect market participants and maintain market stability.

\paragraph{Pausing of Trading}
FINRA states in FINRA Rules\cite{18FINRA2018Rules}, "In the event of any disruption or malfunction in the operation of the electronic communications and trading facilities...a FINRA officer...shall declare as null and void any transaction in a security that occurs after...a regulatory trading halt, suspension or pause..." Similarly, the EU's MiFIR (Markets in Financial Instruments Regulation), (EU) No 600/2014\cite{35EU2014MiFIR}, allows for the temporary suspension of trades under certain conditions with the wording, "..., where the liquidity of that class of financial instrument falls below a specified threshold, temporarily suspend the obligations referred to in Article 8."

These regulations highlight the crucial role of the trading suspension mechanism as a protective measure within financial markets. Designed to maintain market integrity and protect investors during significant volatility, technical malfunctions, or other special circumstances that could impair market functioning, these measures grant regulatory authorities and market operators the authority to temporarily halt trading. By doing so, these provisions aim to prevent panic selling, ensure fair trading practices, and protect the overall stability of the financial system. The ability to suspend trading reflects a preventative approach to risk management, allowing the market to stabilize and be assessed before allowing trading to resume. This emphasizes the absolute necessity of trading suspension mechanisms to maintain orderly market conditions and protect investor interests.

\paragraph{Suspension or Disposal of Contract (kill switch)}
The function commonly referred to as a "kill switch" in smart contracts, which enables pausing or terminating operations, is crucial for controlling operational risks of smart contracts under the principles of a Risk-Based Approach (RBA) by regulatory bodies. The EU, in REGULATION (EU) 2023/2854\cite{34EU2023(EU)2023/2854}, particularly in Article 36 on essential requirements regarding smart contracts for executing data sharing agreements, recommends that smart contracts for data sharing should include "a function that, on the basis of the continued execution of the transactions, allows for the contract to reset, interrupt, or stop operations, particularly to prevent future unintended executions."

These provisions emphasize the absolute necessity of having mechanisms to pause or terminate smart contracts in response to anomalies, risks, or regulatory status changes. The EU's Data Act highlights the importance of resilient access control mechanisms that can prevent unauthorized manipulation, requiring the authority to pause or modify smart contract operations as needed. This recommendation ensures that the token infrastructure can maintain compliance, integrity, and security through preemptive management of smart contracts.

\paragraph{Blacklist Management}
FATF recommends in VIRTUAL ASSETS AND VIRTUAL ASSET SERVICE PROVIDERS\cite{16FATF2021VIRTUAL}, "If a VASP uncovers VA addresses that it has decided not to establish or continue business relations with or transact with due to suspicions of ML\footnote{Money Laundering (ML): The process of circulating proceeds obtained from illegal activities into the legitimate financial system to conceal their origin and convert them into legal assets.}/TF\footnote{Terrorist Financing (TF): The act of providing funds or resources, directly or indirectly, to support terrorist activities.}, the VASP should consider making available its list of 'blacklisted wallet addresses'." Similarly, the SFC includes the technology of "tainted wallet addresses" in the Terms and Conditions for Virtual Asset Trading Platform Operators\cite{30SFC2020Terms}.

These guidelines emphasize the importance of blacklist management in protecting the financial system from risks associated with money laundering and terrorist financing. By requiring virtual asset service providers and financial institutions to maintain and utilize blacklists for suspicious or tainted addresses, the goal is to prevent the flow of illicit funds through the global financial network. The focus on blacklist management reflects a proactive approach to identifying and mitigating risks, demonstrating the necessity of mechanisms to maintain the integrity and stability of the financial markets. Therefore, the practice of blacklist management is absolutely necessary for financial institutions and providers of tokenization and digital asset services to effectively contribute to the global fight against financial crime and enhance the overall security of the financial ecosystem.

\paragraph{Forced Liquidation}
FINRA mentions the necessity of forced liquidation in situations where portfolio margin accounts become insolvent or non-compliant with regulations in FINRA Rules\cite{18FINRA2018Rules}, stating, "A member is required immediately either to liquidate or transfer to another broker-dealer eligible to carry portfolio margin accounts all portfolio margin accounts with positions in related instruments if the member is: (i) insolvent...or (iv) unable to make such computations as may be necessary to establish compliance with such financial responsibility." Similarly, the BIS and IOSCO emphasize the critical role of forced liquidation of a defaulting participant’s positions to manage credit exposure and maintain market stability in Principles for Financial Market Infrastructures\cite{7BIS-IOSCO2012Principles}, stating, "A CCP should have rules and procedures to facilitate the prompt close out or transfer of a defaulting participant’s proprietary and customer positions."

These provisions highlight the absolute necessity of forced liquidation mechanisms within the regulatory framework of capital markets. Through these measures, regulatory bodies and financial institutions can take decisive action in situations where a participant's financial soundness poses a risk to the capital markets or the participant itself, thereby minimizing the possibility of systemic risks. The forced liquidation process emphasizes the importance of maintaining a safe financial environment by proactively managing risks associated with insolvency or regulatory non-compliance, protecting the interests of all market participants.

\subsection{Finality}

Finality is a fundamental attribute that must be maintained to ensure a robust tokenization infrastructure. It necessitates the need for an immutable framework, such as Distributed Ledger Technology (DLT), that ensures transaction records and data integrity cannot be altered, guaranteeing that once transactions are recorded, they cannot be deleted or changed. The principle of immutability reinforced by DLT provides a high level of data integrity by requiring consensus among participants to alter data streams according to specific rules of the distributed ledger. Moreover, finality includes the clarity and certainty of final settlement, ensuring that transactions are irrevocable and unconditional, thereby establishing a trustworthy foundation for asset interoperability. Regulatory bodies that establish technical standards further strengthen finality by stipulating that information related to transactions, including legal documents, be transparent and have clear legal reference, thus protecting the legality and validity of asset transactions. Compliance with finality ensures that the tokenization infrastructure guarantees clear and unambiguous transaction records, significantly contributing to the reduction of disputes and increasing efficiency in the capital markets.

\paragraph{Immutability of the Ledger}
ESMA emphasizes the potential of DLT for tokenized capital markets in its Report on the DLT Pilot Regime On the Call for Evidence on the DLT Pilot Regime and compensatory measures on supervisory data\cite{9ESMA2022Regime}, stating, "data stored on the ledger has a high level of integrity as consensus among participants is necessary to alter data blocks." HKMA further elaborates on this in the Whitepaper 2.0 on Distributed Ledger Technology\cite{20HKMA2017Ledger}, specifying that "blockchains are immutable, once a transaction is written it cannot be erased," highlighting the benefits of finality in demonstrating the ledger's integrity easily. FINRA in FINRA Rules\cite{18FINRA2018Rules} stipulates that "the transaction reports occurred in a DLT cannot be canceled, and it would not be possible to modify records in case of misreporting," thus stating the permanence of transaction records.

These provisions underscore the absolute necessity of record immutability like DLT in the tokenization infrastructure. The requirement for consensus to change data streams, coupled with the immutability of transactions, guarantees a reliable and secure environment for financial transactions. The emphasis by ESMA, HKMA, and FINRA on these aspects demonstrates the crucial role of record immutability in achieving a transparent, trustworthy, and efficient financial system.

\paragraph{Finality of Transactions and Payments}
ISDA emphasizes the importance of having clear mechanisms to resolve disputes, especially in the context of smart derivatives contracts, in LEGAL GUIDELINES FOR SMART DERIVATIVES CONTRACTS: THE ISDA MASTER AGREEMENT\cite{27ISDA2019MASTER}. ISDA suggests, "it will be important for the parties to agree upon a mechanism (whether internal or external to the smart derivatives contract) to determine or verify that any data inputs are correct," highlighting the need for predefined resolution methods to manage discrepancies. Similarly, the FCA in Finalised non-handbook guidance on Cryptoasset Financial Promotions\cite{14FCA2023Finalised} underscores the necessity of clear disclosure regarding changes in crypto asset ownership, stating, "firms should clearly and prominently disclose the changes to legal and beneficial ownership of the crypto asset before a consumer proceeds to enter into a relevant agreement." These provisions emphasize the importance of transparency and clear infrastructure to prevent disputes.

The provisions from ISDA and FCA clearly highlight the importance of implementing more proactive dispute resolution mechanisms within the tokenization infrastructure. Regulatory bodies advocate for systems designed to minimize disputes by establishing clear guidelines for dispute resolution and defining legal frameworks to manage the operation of tokenized capital market infrastructures using DLT.

\paragraph{Attaching Legal Documents}
FINRA specifies the requirements for attaching legal documents during securities transactions in FINRA Rules\cite{18FINRA2018Rules}, stating, "documents required when the laws, regulations, rulings, instructions, or orders of any government...require a license, clearance, certificate, affidavit of ownership, or any similar document...such security shall not be a good delivery unless accompanied by the document or documents so required." This provision emphasizes the importance of complying with legal requirements to ensure the legality and validity of securities transactions. HKMA discusses the innovative application of law to facilitate DLT in Whitepaper 2.0 on Distributed Ledger Technology\cite{20HKMA2017Ledger}, stating, "a digitised version can never receive the same legal standing as its original non-digitised version but it is more a matter of admissibility/weight as evidence in the course of court proceedings," highlighting the challenges and considerations in integrating traditional legal documents into a DLT environment.

The provisions from FINRA and HKMA demonstrate the importance of attaching legal documents as a complement to the contractual legal status within the tokenization infrastructure. Regulatory bodies emphasize the absolute necessity of attaching legal documents within the tokenization infrastructure using DLT to ensure that transactions meet legal standards and regulatory requirements.

\subsection{Tokenizability}
Tokenizability refers to the inherent attributes involved in the design, issuance, and management of digital tokens within a regulatory and technological framework. It encompasses several key aspects, including the ability of tokens to digitally represent assets or ownership, restrictions on transferability to maintain compliance and ensure security, the divisibility or indivisibility of tokens suitable for various financial products, and mechanisms to control token supply. Tokenizability embodies the multifaceted characteristics of digital tokens while recognizing their role as assets with unique properties defined not only by financial instruments but also by the regulatory environment, technological infrastructure, and intended use cases. Tokenizability accommodates the complexities of issuing and operating digital tokens and highlights the need for robust and adaptable functionalities for utility and financial product guidelines within the broader financial ecosystem.

\paragraph{Token Expired Time and Token Transfer Restrictions}
One of the most crucial aspects of tokenization infrastructure is ensuring regulated control over the transferability and expiration of tokens, which is essential for maintaining the unique guidelines of financial products. FINRA, in its FINRA Rules\cite{18FINRA2018Rules}, specifically under "2360. Options," states, "The term 'expiration date' of an option contract...means the day and time fixed in accordance with the rules of The Options Clearing Corporation for the expiration of such option contracts...." This provision emphasizes the control of the product lifecycle according to clear and predefined expiration parameters for option contracts, which is essential for the orderly functioning of the options market and prevention of fraud and manipulation. Additionally, ESMA in the Consultation Paper, On the draft Guidelines on the conditions and criteria for the qualification of crypto-assets as financial instruments\cite{11ESMA2024instruments}, under "6.3 Annex II - Draft Guidelines on the classification of crypto-assets as financial instruments," details "transfer restrictions," stating, "A crypto-asset can be designed in a way that it does not allow for any transfer in capital markets." This provision is crucial for maintaining trust in the unique properties of financial products that do not allow holders to transfer or sell to anyone other than the issuer, according to financial product guidelines, preventing fraudulent transactions and ensuring that all market participants are aware of and can safely engage with the product's unique attributes.

These provisions comprehensively underscore the absolute necessity of structural control over the expiration and transferability of tokens according to the uniqueness of financial products. Such regulations ensure that tokenized assets adhere to the same rigorous guidelines as traditional financial products, protecting investors and maintaining a fair and orderly market.

\paragraph{Issuance of Tokenized Cash and Issuance of Tokenized Securities}
The process of tokenization in DLT, which digitally represents assets, is another fundamental function within the token infrastructure that ensures effective and complete digitization across securities and digital cash that can settle securities. IOSCO, in Financial Technologies (Fintech)\cite{24IOSCO2017Technologies}, especially in the "Distributed Ledger Technology (DLT)" section, highlights the importance of tokenization by stating, "A “token” represents an asset or ownership of an asset. Such assets can be currencies, commodities or securities or properties." This statement emphasizes that tokenized assets can represent not only securities and commodities but also currencies, suggesting that tokens transformed into digital format can represent any form of asset and ownership of assets. Additionally, ESMA in the Consultation Paper, On the draft Guidelines on the conditions and criteria for the qualification of crypto-assets as financial instruments\cite{11ESMA2024instruments}, particularly in "Annex II - Draft Guidelines on the classification of crypto-assets as financial instruments," describes the issuance of tokenized securities or non-fungible tokens (NFTs), stating "to be unique, NFTs should be considered distinct and irreplaceable where their characteristics and/or the rights they provide are not identical to the other crypto-assets issued by the same (or any other) issuer." This provision highlights the uniqueness and irreplaceability of certain tokenized assets in the digital asset space, emphasizing the importance of utilizing NFTs for investment and asset management due to their ability to express unique value and ownership.

These provisions from IOSCO and ESMA show the necessity of issuing various foundational token properties within the tokenized asset ecosystem. Tokenization facilitates the digital representation of a wide range of assets, enhancing liquidity and marketability, and provides an innovative way to manage and invest assets within a safe and regulated framework, essential for the evolution and expansion of the digital economy.

\paragraph{Controlling Transactions Involving Splitting Below Decimal Units and Token Burning}
The functionality to divide tokens into smaller units and mechanisms for token burning (or removing from circulation) are critical features within the tokenization infrastructure, directly impacting the liquidity, market efficiency, and value stability of digital assets. While FINRA, in FINRA Rules\cite{18FINRA2018Rules}, specifically under "4210. Margin Requirements (g) Portfolio Margin," specifies requirements for liquidation or transfer in cases of insolvency or regulatory non-compliance, it does not directly quote any specific provisions regarding token burning. However, the context of managing portfolio risk and position liquidation requirements can be analogous to the importance of controlling token supply through burning mechanisms. This process is essential for maintaining compliance with financial product guidelines.

Furthermore, discussions on token divisibility imply the importance of token divisibility for financial products and the digital economy. Divisibility is crucial for ensuring access to and use of digital assets across a range of investment sizes, thereby enhancing utility and participation in the broader financial ecosystem.

It is clear that both the divisibility and burning mechanisms of tokens are indispensable for a robust token ecosystem. The feature of token divisibility, emphasizing the efficiency of financial products, is determined by considering market demand, potential conflicts with existing regulations, and operational requirements comprehensively.

\paragraph{Gasless Support and Asset Class Management}
In the evolving digital asset environment, the concept of allowing a third party to pay for blockchain gas fees, known as meta transactions (ERC-2771)ERC-2771: Secure Protocol for Native Meta Transactions\cite{4Person2020ERC-2771}, and asset class management (management by type of financial product) emerges as the cornerstone of infrastructure that can manage and expand various tokenized asset ecosystems. Although the regulatory agency's regulations do not provide specific citations on Gas sponsorship, the concept of gasless transactions represents a significant advancement in user accessibility and efficiency, reducing transaction barriers in the blockchain network by allowing for the handling of transaction fees. This function, provided not by the users themselves but by the operators of the tokenization infrastructure, is crucial in enhancing the usefulness and inclusiveness of the digital asset system.

In IOSCO's Policy Recommendations for Crypto and Digital Asset Markets\cite{22IOSCO2023Policy}, it is advised that "CASP maintain accurate and up-to-date records and accounts of Client Assets that readily establish the precise nature, amount, location, and ownership status of Client Assets and the clients for whom the assets are held." This emphasizes the importance of asset class management. Asset class management can systematize the token supply management according to the dynamic requirements of the regulatory environment for each financial product by adopting the token issuance form that can be structured from financial products like the Token Taxonomy Framework (TTF)\footnote{Token Taxonomy Framework (TTF): A token classification framework that enables collaboration in modeling financial products and defining new business models to bridge the gap between developers and regulatory agencies} of the Inter Work Alliance (IWA). According to TOKEN TAXONOMY: The Need for Open-Source Standards Around Digital Assets\cite{5IWA2020TAXONOMY} by The TTF, in the case of bonds, they can be denoted as
\[
tF\{\sim d,t,c\}
\]
This means that bonds are tF (Fungible token) and have an asset class structure characterized by $\sim$d (Non-Subdivisible), t (Transferable), c (Compliant). Specifically, ‘Non-Subdivisible’ ($\sim$d) is the negation of the TTF Divisible behavior, which the TTF defines as "An ability for the token to be divided from a single whole token into fractions, which are represented as decimal places," so that a Non-Subdivisible token cannot be split below a single whole unit. ‘Transferable’ means "The Transferable behavior provides the owner the ability to transfer the ownership to another party or account." ‘Compliant’ indicates that "A regulated token needs to comply with several legal requirements, especially KYC and AML."

It becomes evident that asset class management facilitated by the IOSCO and IWA's TTF initiative is crucial for asset management efficiency and adapting to changing regulations. The tokenization infrastructure, through Gas sponsorship, can promote easier access to transactions and provide a framework that ensures strict management of asset classes according to various financial product guidelines, thereby facilitating the mass adoption of tokenized capital markets.

\section{Methods}
RCP includes all comprehensive regulatory compliance requirements across the tokenization process, trading, and overall tokenization infrastructure as examined in the previous chapter, similar to security token protocols like ERC-1400 and ERC-3643, where the procedure is carried out. However, RCP exceeds the regulatory compliance requirements of other security token protocols, eliminating regulatory uncertainty throughout the entire process of asset tokenization, trading, and redemption. Focusing on RCP-based tokenization services, it aims to explain RCP through each procedure and pseudocode in three scenarios: 1) Bond Issuance and Lifecycle Management Scenario, 2) Carbon Credit Scenario, 3) Interoperability Scenario between TradFi and DeFi.

\subsection{Bond Issuance and Lifecycle Management Scenario}

This scenario represents the entire process of tokenizing traditional financial assets, specifically bonds, from preparation for issuance through to issuance, trading, and finally to maturity and settlement. It covers the entire process of bond issuance and lifecycle management, explaining the roles and interactions of various participants including issuers, legal counsel, tokenization services, brokers, KYC, investors, and regulatory authorities. RCP acts as a key element in this process, ensuring regulatory compliance while providing transparency and reliability. Specifically, through various regulatory compliance controls of RCP such as customer identity verification, contract version tracking, token expired time and Controlling Transactions Involving Splitting Below Decimal Units, and Asset Class Management, the issuance process's security and efficiency are guaranteed, thereby enhancing the safety and regulatory compliance of the financial market. This scenario demonstrates how RCP, unlike ERC-1400 and ERC-3643, effectively adheres to the recommendations and guidelines of regulatory authorities in the asset tokenization process.

\begin{figure*}[ht]

  \centering
  \includegraphics[scale=0.45]{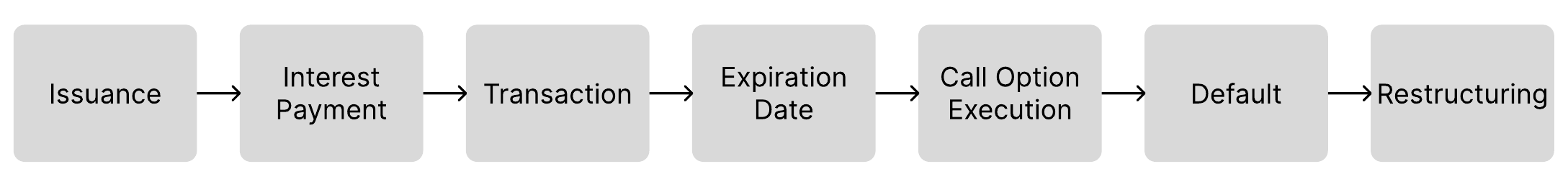}
  \caption{Flow of Bond Issuance and Lifecycle Management Scenario}
  \label{fig:fig1}

\end{figure*}

In the financial technology domain, the RCP is pivotal for bond tokenization and lifecycle management. Utilizing Distributed Ledger Technology (DLT) and smart contracts, the RCP enforces compliance, ensures transparency, and secures operations. The following sections detail this process, including the critical role of legal counsel in notarization, which is integral to the legal and regulatory compliance checks.

The initial stage involves legal and regulatory compliance checks by issuers and their legal counsel, including notarization to ensure the authenticity and enforceability of the documents. This is represented as:
\[
\Gamma_{\text{prep}} = \sum_{\omega \in \Omega} \rho(\omega, \lambda_{\text{legal}}, \sigma_{\text{RCP}}, \nu_{\text{notarization}})
\]
where $\Gamma_{\text{prep}}$ indicates preparatory operations, $\Omega$ the set of requirements, $\rho$ the compliance function, $\lambda_{\text{legal}}$ legal advisories, $\sigma_{\text{RCP}}$ the RCP's compliance mechanisms, and $\nu_{\text{notarization}}$ the notarization process by legal counsel.

The next step, tokenization and issuance, involves using smart contracts to either create Tokenized Cash (FT) or Securities (NFT), within the standards set by \(\Delta_{\text{RCP}}\), with legal counsel providing notarization to ensure the contracts' legal validity. This process can be mathematically represented as:
\[
\Phi_{\text{token}} = \Delta_{\text{RCP}} \cap (\Theta_{\text{FT}} \oplus \Theta_{\text{NFT}}) \cap \nu_{\text{notarization}}
\]
Here, \(\Phi_{\text{token}}\) denotes the tokenization operations, \(\Delta_{\text{RCP}}\) represents the RCP standards, \(\Theta_{\text{FT}}\) and \(\Theta_{\text{NFT}}\) indicate the types of tokens that can be created, and \(\oplus\) symbolizes the exclusive OR (XOR) operation, highlighting that either FT or NFT can be chosen for tokenization within the RCP standards framework.

Ensuring market integrity involves setting up KYC and trading restrictions, with legal counsel's notarization ensuring the compliance of these processes with regulatory standards. This is modeled as:
\[
\Lambda_{\text{KYC}} = \xi_{\text{RCP}}(\kappa_{\text{KYC}}, \tau_{\text{restrict}}, \nu_{\text{notarization}})
\]
$\Lambda_{\text{KYC}}$ denotes KYC and trading restrictions setup, $\xi_{\text{RCP}}$ the RCP's management function, $\kappa_{\text{KYC}}$ the KYC procedures, $\tau_{\text{restrict}}$ the trading restrictions, and $\nu_{\text{notarization}}$ the notarization process ensuring regulatory compliance.

Secure and compliant transactions are facilitated in the trading and compliance phases, with notarization playing a role in the verification of compliance documents and agreements. Described by:
\[
\Omega_{\text{trade}} = \eta_{\text{RCP}}(\mu_{\text{trade}}, \nu_{\text{compliance}}, \nu_{\text{notarization}})
\]
$\Omega_{\text{trade}}$ represents trading and compliance operations, $\eta_{\text{RCP}}$ the RCP's function, $\mu_{\text{trade}}$ trade requests, $\nu_{\text{compliance}}$ compliance checks, and $\nu_{\text{notarization}}$ the notarization of compliance documents.

The process concludes with maturity and settlement, where assets are transferred following gasless settlements, and notarization ensures the legal validity of the settlement documents and agreements:
\[
\Xi_{\text{settle}} = \zeta_{\text{RCP}}(\alpha_{\text{maturity}}, \beta_{\text{settlement}}, \nu_{\text{notarization}})
\]
$\Xi_{\text{settle}}$ indicates maturity and settlement operations, $\zeta_{\text{RCP}}$ the settlement function, $\alpha_{\text{maturity}}$ maturity checks, $\beta_{\text{settlement}}$ settlement executions, and $\nu_{\text{notarization}}$ the notarization process ensuring the legal validity of settlement documents and agreements.



\begin{algorithm}
    \caption{Bond Issuance and Lifecycle Management}
    \begin{algorithmic}
    \State \textbf{Preparation Phase:}
    \If{Legal and Regulatory Compliance met}
        \State Proceed to Tokenization and Issuance
    \Else
        \State Halt and Review Requirements
    \EndIf
    
    \State \textbf{Tokenization and Issuance Phase:}
    \State Define and Deploy Smart Contracts
    \State Issue Tokenized Cash (FT) and Securities (NFT)
    \State Set Regulatory Compliance and Trading Restrictions
    
    \State \textbf{KYC and Trading Restrictions Setup:}
    \For{each investor}
        \If{KYC Approved}
            \State Set Trading Restrictions
        \Else
            \State Request Additional Information
        \EndIf
    \EndFor
    
    \State \textbf{Market Trading Phase:}
    \While{Market Open}
        \If{Trade Request Complies with Restrictions}
            \State Execute Trade
        \Else
            \State Reject Trade
        \EndIf
    \EndWhile
    
    \State \textbf{Maturity and Settlement Phase:}
    \If{Bond Maturity Reached}
        \State Prepare for Settlement
        \State Calculate Principal and Interest
        \State Execute Gasless Settlement
        \State Transfer Assets to Investors
        \State Record Settlement for Audit
    \EndIf
    
    \State \textbf{Auditing and Reporting Phase:}
    \State Perform Real-time Transaction Monitoring
    \State Maintain Record Immutability
    \State Automated Regulatory Reporting
    
    \end{algorithmic}
    \end{algorithm}


In the \textbf{Preparation Phase}, establishing a robust framework of legal and regulatory compliance is paramount. The formulation is:
\[
\Upsilon_{\text{prep}} = \bigcup_{\lambda \in \Lambda} \sigma(\lambda) \times \bigcap_{\delta \in \Delta} \varphi(\delta)
\]
$\Upsilon_{\text{prep}}$ symbolizes preparatory operations, $\Lambda$ represents legal advisories, $\sigma$ maps legal advisories to their compliance metrics, $\Delta$ is the set of regulatory requirements, and $\varphi$ verifies compliance for each requirement. This captures the alignment of legal advisories with regulatory requirements.

Proceeding to the \textbf{Tokenization and Issuance Phase}, smart contracts facilitate the tokenization process. The framework is given by:
\[
\Omega_{\text{token}} = \sum_{t \in \mathcal{T}} \psi(t, \mathcal{S})
\]
$\Omega_{\text{token}}$ denotes tokenization operations, $\mathcal{T}$ the period of execution, $\psi$ the tokenization function dependent on $\mathcal{S}$, the classification of tokens (FT and NFT). This integral illustrates the process of token issuance and management.

The \textbf{KYC and Trading Restrictions Setup Phase} introduces measures for investor scrutiny and transactional oversight. The operations are described by:
\[
\Theta_{\text{KYC}} = \sum_{i=1}^{n} \kappa(i) \odot \tau(i)
\]
$\Theta_{\text{KYC}}$ involves setting up KYC and trading restrictions, $\kappa(i)$ is the KYC verification function for each investor, $\tau(i)$ the trading restriction function, and $\odot$ the Hadamard product, applying trading restrictions based on KYC outcomes.

The \textbf{Market Trading Phase} enforces compliance and integrity through regulatory checks. The operations are detailed by:
\[
\Phi_{\text{trade}} = \bigoplus_{j \in J} \rho(j) \otimes \mu(j)
\]
$\Phi_{\text{trade}}$ covers market trading operations, $J$ the set of trade requests, $\rho(j)$ the compliance check for each trade, $\mu(j)$ the market execution function, $\bigoplus$ the direct sum, and $\otimes$ the tensor product, showing the interaction between compliance checks and market execution.

The discussion concludes with the \textbf{Maturity and Settlement Phase}, where the settlement process is outlined by:
\[
\Psi_{\text{settle}} = \bigvee_{k in K} \alpha(k) \wedge \beta(k)
\]
$\Psi_{\text{settle}}$ represents settlement operations, $K$ the set of matured bonds, $\alpha(k)$ the maturity verification function, $\beta(k)$ the settlement execution function, $\bigvee$ the logical OR, and $\wedge$ the logical AND, integrating maturity verification and settlement execution.

\subsection{Carbon Credit Scenario}

In this scenario, we examine the application of RCP focusing on the tokenization process of carbon credits. It describes how various participants such as issuers, investors, and regulatory bodies interact with each other, and how RCP provides differentiated regulatory compliance features compared to existing protocols like ERC-3643 and ERC-1400.

RCP enhances regulatory compliance throughout the entire process of carbon credit tokenization, particularly through Attaching Legal Documents, role-based permission settings, and the setting of token expiration and transfer restrictions. It is designed to thoroughly meet the requirements of regulatory bodies, while also increasing the flexibility of tokenized assets through the setting of token divisibility and asset class management, thereby managing the complexity of regulatory compliance.

This functional superiority makes RCP a preferred choice over existing protocols for the tokenization of specific assets like carbon credits. The introduction of RCP enables efficient management and trading of carbon credits, enhances market transparency, and ensures regulatory compliance. This underscores the importance of RCP in the asset tokenization field and suggests its future role in the capital markets.

\begin{figure*}[ht]

  \centering
  \includegraphics[scale=0.45]{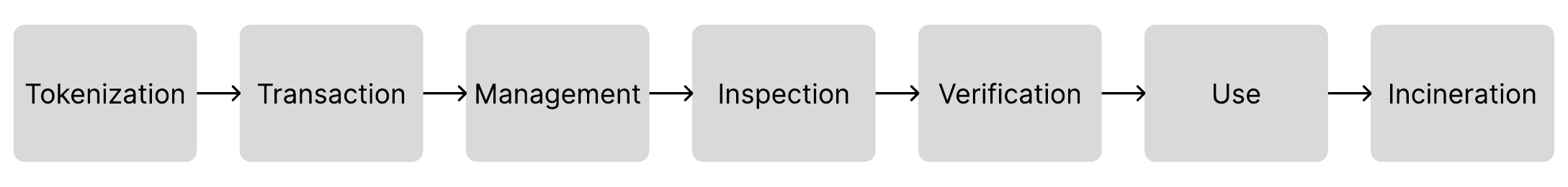}
  \caption{Flow of Carbon Credit Scenario}
  \label{fig:fig1}

\end{figure*}

The tokenization process of carbon credit is designed based on the RCP to meet the complex regulatory environment and technical requirements. This process starts with the issuer attaching legal documents and setting role-based permissions ($\mathcal{F}_{\text{prep}}$), followed by setting the token's validity period and transfer restrictions through RCP ($\mathcal{F}_{\text{config}}$). Token issuance is implemented as Non-Fungible Tokens (NFTs) ($\mathcal{F}_{\text{NFT}}$) with asset class management ($\mathcal{F}_{\text{class}}$), and it sets options for token splitting and burning ($\mathcal{F}_{\text{split}}, \mathcal{F}_{\text{burn}}$). In the regulatory compliance verification process, Customer Identity Verification ($\mathcal{F}_{\text{KYC}}$), Contract Version Tracking ($\mathcal{F}_{\text{track}}$), and Blacklist Management ($\mathcal{F}_{\text{blacklist}}$) play crucial roles. Through these processes, RCP ensures regulatory compliance throughout the entire carbon credit tokenization process, providing transparency and reliability as a key element. The RCP-based approach to carbon credit tokenization presented in this study offers a method that more thoroughly complies with the recommendations and guidelines of regulatory bodies compared to existing protocols, ensuring the security and efficiency of the asset tokenization process and enhancing the safety and regulatory compliance of the financial market.

In the trading and management process, RCP plays a key role in ensuring interoperability between traditional financial markets and decentralized financial markets. At this stage, requests for the purchase, sale, or exchange of carbon credit tokens ($\mathcal{G}{\text{request}}$) are transmitted to RCP through exchanges, and RCP verifies the restrictions and regulations of the transaction ($\mathcal{G}{\text{verify}}$). This process can include requests for asset freezing, recovery ($\mathcal{G}{\text{freeze}}, \mathcal{G}{\text{recover}}$), and token splitting, burning ($\mathcal{G}{\text{split}}, \mathcal{G}{\text{burn}}$). Regulatory bodies monitor transactions and asset management ($\mathcal{G}{\text{monitor}}$) and can request the suspension of transactions or financial products ($\mathcal{G}{\text{suspend}}$) if necessary. These interactions are crucial for RCP to continuously ensure regulatory compliance, maintaining market transparency and reliability. The RCP-based approach presented in this study strengthens regulatory compliance in the trading and management process, enabling safe trading and management of tokenized assets.

The audit and verification stage is a critical part of the RCP, playing a vital role in securing regulatory compliance and security throughout the carbon credit tokenization process. In this stage, interactions among issuers, investors, and regulatory bodies verify the validity and compliance status of the tokens. RCP verifies all transactions meet the latest regulatory requirements by confirming customer identity through $\mathcal{F}_{\text{KYC}}$, tracking contract versions with $\mathcal{F}_{\text{track}}$, and restricting transactions of entities on the blacklist using $\mathcal{F}_{\text{blacklist}}$, which is essential for ensuring the reliability of tokenized assets. Audit institutions, based on information provided by RCP, compile audit results and report them to regulatory bodies, thereby enhancing the transparency and regulatory compliance of the entire carbon credit tokenization process. Through these processes, RCP offers a more thorough regulatory compliance verification mechanism compared to existing protocols (ERC-3643, ERC-1400), ensuring the reliability and safety of carbon credit tokenization.

The process of using and burning carbon credit utilizes the core functionalities of the RCP. In this process, consumers submit requests to RCP for using or burning carbon credit, and RCP verifies these requests ($\mathcal{F}_{\text{verify}}$). For use requests, RCP checks the validity using the 'transfer restrictions' feature and, if approved, proceeds with the use approval and ownership transfer process ($\mathcal{F}_{\text{use}}$). Subsequently, information on the used carbon credit is reported to regulatory bodies, ensuring regulatory compliance through the attachment of legal documents. In the case of carbon credit burn requests, RCP verifies the request and, upon approval, proceeds with the burn approval and ensures record immutability ($\mathcal{F}_{\text{burn}}$). The process and legal compliance of the burn are reported to audit institutions, which then report to regulatory bodies, thereby enhancing the transparency and regulatory compliance of the use and burn process of carbon credit. Through these processes, RCP manages the use and burning of carbon credit in compliance with regulatory standards, ensuring the reliability and safety of carbon credit tokenization.

\begin{algorithm}
    \caption{Carbon Credit Tokenization and Transaction Management}
    \begin{algorithmic}[1]

        \If{Preparation for Issuance Completed}
            \State Attach Legal Documents and Compliance
            \State Role-Based Permission Setting
            \State Setting Token Validity Period
            \State Setting Transfer Restrictions
        \Else
            \State Check Preparation Status
        \EndIf

        \If{Investor Requests Purchase, Sale, or Exchange of Carbon Credit Tokens}
            \State Request Transmitted to RCP via Exchange
            \If{RCP Reviews Transaction Restrictions and Regulations}
                \State Transaction Approval and Recording
            \Else
                \State Transaction Rejection and Reason Notification
            \EndIf
        \EndIf
        \If{Issuer Requests Asset Freeze or Recovery}
            \State RCP Approves Request and Records
        \Else
            \If{Investor Requests Token Split or Burn}
                \State RCP Approves Request and Records
            \EndIf
        \EndIf

        \If{Issuer Requests Asset Freeze or Recovery}
        \State RCP Approves Request and Records
    \Else
        \If{Audit and Verification Request Exists}
            \State Receive Request from Audit Institution
            \If{Role-Based Permission Setting Verification}
                \State Provide Role-Based Permission Setting Information
            \EndIf
            \If{Legal Document Attachment and Compliance Verification}
                \State Provide Legal Document and Compliance Information
            \EndIf
            \If{Transaction Records and Activity Logs Request}
                \State Provide Transaction Records and Activity Logs
            \EndIf
            \If{Customer Identity Verification and Transaction Restriction Verification}
                \State Provide Verification Results and Related Information
            \EndIf
            \If{Asset Freeze and Blacklist Management Verification}
                \State Provide Verification Results and Related Information
            \EndIf
            \If{Asset Recovery and Forced Liquidation (Burn) Process Verification}
                \State Provide Process Verification Results and Related Information
            \EndIf
        \EndIf
    \EndIf

            \end{algorithmic}
        \end{algorithm}
        \begin{algorithm}
        \ContinuedFloat
        \caption{Carbon Credit Tokenization and Transaction Management(continued)}
        \begin{algorithmic}

    \If{Consumer Requests Use of Carbon Credit Rights}
        \State RCP Verifies Request ('Using Transfer Restrictions')
        \If{Request Approved}
            \State Ownership Transfer and Use Approval
            \State Reporting Use to Regulatory Body and Attaching Legal Documents
        \Else
            \State Use Request Rejection and Reason Notification
        \EndIf
    \ElsIf{Consumer Requests Burn of Carbon Credit Rights}
        \State RCP Verifies Burn Request
        \If{Request Approved}
            \State Burn Approval and Ensuring Record Immutability
            \State Reporting Burn Process and Legal Compliance to Audit Institution
        \Else
            \State Burn Request Rejection and Reason Notification
        \EndIf
    \EndIf

    \end{algorithmic}
\end{algorithm}

The carbon credit tokenization process begins with the issuer accessing the tokenization service to convert carbon credit into digital assets. In this process, the issuer goes through steps such as attaching legal documents and compliance, setting role-based permissions, setting the token's validity period, and setting transfer restrictions. These steps are performed using the RCP's compliance functions.

Once the token issuance is complete, RCP provides the issuer with token issuance confirmation and proceeds with the regulatory compliance verification process with regulatory bodies. Regulatory bodies review the information provided through RCP to confirm compliance and notify RCP.

Carbon credit tokens are supplied to the market and traded through exchanges. During the trading process, RCP checks whether the transfer and trade of tokens meet regulatory compliance requirements.

Audit institutions collaborate with RCP to conduct audits on the entire carbon credit tokenization process, and the audit results are guaranteed record immutability through digital certification.

Consumers can request the use of carbon credit through RCP, and RCP verifies the request before proceeding with use approval and ownership transfer. Consumers can also request the burning of carbon credit, and RCP verifies and approves the burn request. This process also involves regulatory compliance reporting to regulatory bodies.

Through these processes, carbon credit tokenization is efficiently managed based on regulatory compliance, enhancing market transparency and reliability.

{\small
\[
\text{Tokenization process} = \sum_{i=\text{issuance}}^{\text{trade}} \text{Regulatory Compliance Function}(i)
\]
}
\begin{multline*}
{\small \text{Regulatory Body Verification Function}} = \\ {\small
f(\text{Compliance Information}) = \begin{cases} 
\text{Approval}, & \text{if info = compliant} \\
\text{Rejection}, & \text{otherwise} 
\end{cases}
}
\end{multline*}

In the carbon credit tokenization process, trading and management are key functions of RCP. This process includes various stages from token issuance to trading, and ultimately to use or burning. Especially in the trading and management phase, it is important to verify that the transfer of tokens meets regulatory compliance requirements. To this end, RCP uses the following mathematical model to define the trading and management process.

\begin{multline*}
F_{\text{Trading and Management}} = \bigcup_{i=1}^{n} f_{\text{Regulatory Compliance Verification}}(T_i) \\ \oplus f_{\text{Transaction Execution}}(T_i) \oplus f_{\text{Audit and Verification}}(T_i)
\end{multline*}

Here, $F_{\text{Trading and Management}}$ represents the entire process of trading and management, and $T_i$ represents individual token transactions. $f_{\text{Regulatory Compliance Verification}}$ is a function that verifies each transaction meets regulatory compliance requirements, $f_{\text{Transaction Execution}}$ is a function that executes the actual token transactions. Finally, $f_{\text{Audit and Verification}}$ is a function that ensures transactions are accurately recorded and meet the audit requirements of regulatory bodies.

The audit and verification stage is a critical phase in ensuring regulatory compliance and security throughout the carbon credit tokenization process. In this stage, the accuracy and compliance status of the information provided through RCP are verified. The audit and verification function can be defined as follows:

\begin{multline*}
F_{\text{Audit and Verification}} = \\ \sum_{j=1}^{m} \left( f_{\text{Information Verification}}(I_j) + f_{\text{Compliance Confirmation}}(I_j) \right)
\end{multline*}

Here, $F_{\text{Audit and Verification}}$ represents the entire audit and verification process, and $I_j$ represents audit target information items. $f_{\text{Information Verification}}$ is a function that verifies the accuracy of the provided information, and $f_{\text{Compliance Confirmation}}$ is a function that confirms whether the information meets regulatory compliance requirements.

The audit and verification process plays an important role in ensuring the transparency and reliability of RCP.

The process of using and burning carbon credit is one of the core elements of carbon credit tokenization and is managed through the RCP. This process is explained through mathematical models and pseudocode.

\begin{multline*}
F_{\text{Use and Burn}} = \sum_{k=1}^{p} \Bigl( f_{\text{Use Verification}}(U_k) + f_{\text{Burn Verification}}(U_k) \\ + f_{\text{Reporting and Audit}}(U_k) \Bigr)
\end{multline*}

Here, $F_{\text{Use and Burn}}$ represents the process of using and burning carbon credit, and $U_k$ represents individual use or burn requests. $f_{\text{Use Request Verification}}$ is a function that verifies whether a use request meets regulatory compliance requirements, $f_{\text{Burn Request Verification}}$ is a function that verifies whether a burn request meets regulatory compliance requirements. Lastly, $f_{\text{Reporting and Audit}}$ is a function that ensures the request processing is accurately recorded and meets the audit requirements of regulatory bodies.

\subsection{Interoperability Scenario between TradFi and DeFi}

 This scenario examines the process of interoperability between tokenized assets in traditional finance and DeFi platforms, focusing on the application of DLT-based DAML to fully comply with the RCP, which is difficult to meet at the ERC protocol level. This scenario involves contract modeling of traditional financial assets including rights and obligations through DAML, implementing traceable privacy through DAML to meet the privacy and traceability of RCP, and providing bidirectional interoperability services between DLT and blockchain through the oraclizer service to comply with the completeness of RCP through an atomic processing process across settlement and clearing. This presents a solution that simultaneously satisfies smooth asset interoperability between traditional financial institutions and DeFi platforms and the recommendations of high-level regulatory bodies.

\begin{figure*}[ht]

  \centering
  \includegraphics[scale=0.45]{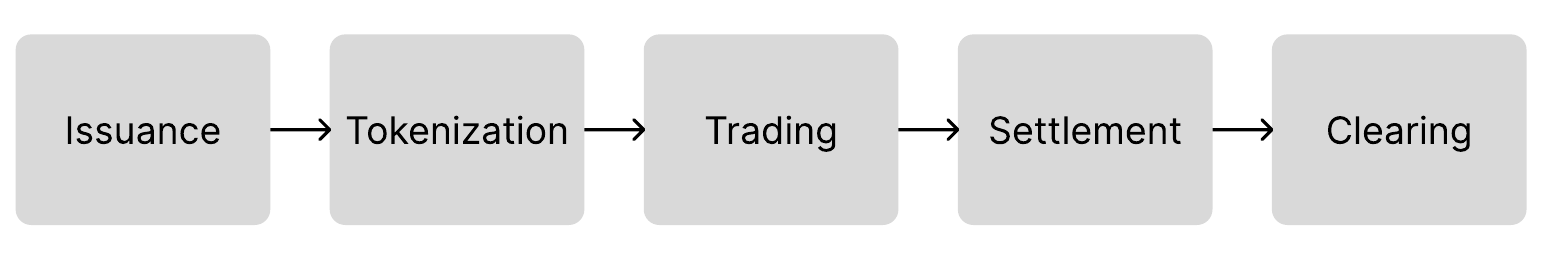}
  \caption{Flow of Interoperability Scenario between TradFi and DeFi}
  \label{fig:fig1}

\end{figure*}


The bond issuance process in traditional financial institutions consists of three stages: checking regulatory compliance requirements, bond information modeling, and applying the RCP. In this process, the RCP provides a framework for compliance, while DAML plays a crucial role in the implementation of bond information modeling and RCPs. Compared to ERC-3643 and ERC-1400, RCP offers more comprehensive regulatory compliance functionalities, including customer identity verification, asset freeze and recovery, transaction restrictions, and limit settings. This ensures that the trading of tokenized assets complies meticulously with the regulations and guidelines of global financial regulatory bodies. However, RCP alone has limitations in efficiently modeling and executing complex contract logic. To overcome these limitations, DLTs like DAML can be utilized. DAML abstracts the rights and obligations of contracts, and RCP allows for detailed control over asset freeze, recovery, and transaction restrictions through role-based permission settings. Additionally, DAML provides functionalities like strong privacy compliance based on a 'need to know' basis, where uninvited users cannot see any contract content that includes specific transaction information, and integrated time management, managing time-based conditions such as the validity period of contracts.

To elucidate this process insightfully, the regulatory compliance requirements in the bond issuance process can be represented as \(C_{reg}\). Here, \(C_{reg}\) is defined by the RCP \(P_{RCP}\), interacting with the bond information \(B_{info}\) modeled using DAML. This simplifies the management of complex regulatory environments conceptually.

\[
C_{reg} = f(P_{RCP}, B_{info})
\]

Here, \(f\) represents the process of meeting regulatory compliance requirements. This interaction ensures regulatory compliance in the bond issuance process and enables the execution of contract logic implemented through DAML.

The tokenization process is a key step in enabling interoperability between traditional financial assets and DeFi assets. In this process, the role of RCP is to ensure regulatory compliance, while DAML provides essential tools for implementing these protocols. Through the process of tokenization services, we quantify the complexity of this process and highlight the complementary functionalities of RCP and DAML.

In the tokenization process, we define the issuer \(P\), the tokenized asset \(A\), and the token \(T\). The issuer \(P\) executes the process \(f_{tokenize}: (P, A) \rightarrow T\), converting asset \(A\) into token \(T\). This process is regulated by RCP and implemented using DAML.

\[
T = f_{tokenize}(P, A)
\]

Here, \(T\) represents the tokenized asset, created according to the regulatory compliance requirements of RCP and the contract logic defined by DAML. This process ensures the regulatory compliance of the tokenization process and emphasizes the necessity of implementation through DAML.

The trading scenario on the DeFi platform can be explained by utilizing RCP and DAML to handle the trading of tokenized assets in a regulatory-compliant and efficient manner. In this process, RCP is responsible for regulatory compliance verification, while DAML manages contract execution and atomic trade processing.

Particularly, the process of ensuring the atomicity of transactions can be understood as follows:

\[
{\scriptsize	\text{Let } T = \text{Transaction}, \; C = \text{Contract Conditions}, \; V = \text{Verification by RCP}}
\]
\[
\text{Atomicity}(T, C, V) = 
\begin{cases} 
  \text{Execute}(T) & \text{if } V(C) = \text{True} \\
  \text{Abort}(T) & \text{otherwise}
\end{cases}
\]

Here, `Execute(T)` occurs when transaction `T` satisfies all contract conditions `C` and regulatory compliance verification `V` by RCP. Otherwise, the transaction is aborted with `Abort(T)`. This demonstrates how the contract logic of DAML and the regulatory compliance verification by RCP work together to ensure the safety and completeness of transactions.

The settlement and clearing process is a key step in ensuring the finality of transactions on the DeFi platform. In this process, RCP and DAML are responsible for regulatory compliance and efficient execution of contract logic, respectively, and their interaction can be explained through the sequence diagram.

To describe this interaction, the settlement and clearing process can be represented as the following function:

\[
F_{settlement} = f(RCP_{compliance}, DAML_{logic})
\]

Here, \(F_{settlement}\) represents the settlement and clearing process, \(RCP_{compliance}\) denotes the verification of regulatory compliance by RCP, and \(DAML_{logic}\) represents the execution of contract logic implemented using DAML. This function illustrates how the regulatory compliance framework of RCP and the contract modeling and execution capabilities of DAML work together to ensure the safety and finality of transactions.

In conclusion, the regulatory compliance framework of RCP and the contract modeling and execution capabilities of DAML operate in a complementary manner to build a robust system that can safely and efficiently handle every stage of financial transactions, from bond issuance by traditional financial institutions to tokenization, trading, and settlement and clearing on DeFi platforms. Through this process, the digitalization and innovation of the financial system are promoted, enhancing transparency and trust in the global financial market.

\section{Discussions}

\subsection{Comparison}
RCP stands as the underlying protocol for executable protocols within tokenized capital markets, establishing a standard that adheres to all regulatory guidelines and features pertinent to tokenized assets. To assess where existing standards stand against this regulatory reference, we conducted a thorough examination of the ERC-20 standard, a cornerstone in the DeFi ecosystem, together with ERC-7943 (the recently finalized uRWA interface)\cite{37LoBuglio2025ERC7943} and ERC-1400 and ERC-3643, which are purpose-built for the tokenization of assets. RCP functions here as a value-neutral benchmark rather than a competing standard: it records what each standard's base specification can express at the token level, and where a given regulatory requirement is inherently off-chain and therefore outside the reach of any token standard. Certain requirements, notably privacy and traceability, pose compliance challenges at the ERC protocol level and require the support of DLTs like DAML, so the comparison focuses on the on-chain-addressable subset of requirements.

For interoperability with tokenized assets in traditional finance, it's crucial to fully satisfy regulatory recommendations and control functions, which is challenging at the ERC protocol level. Particularly, aspects like privacy and traceability require the support of DLTs like DAML. We therefore restrict the comparison to the portion of the requirements that a token standard can address on-chain, setting aside the parts that DLT handles. We organized Table 2 to intuitively compare how the existing protocols ERC-20, ERC-7943, ERC-1400, and ERC-3643 meet the regulatory recommendations, based on the EIP documents proposed for each standard. This table allows us to see, requirement by requirement, where each standard provides a base-specification mechanism and where the requirement is inherently off-chain.


\begin{table*}[t]
  \centering
  \begin{tabular}{lC{1.6cm}C{1.6cm}C{1.6cm}C{1.6cm}}
    \toprule
    RCP & ERC-20 & ERC-7943 & ERC-1400 & ERC-3643 \\
    \midrule
    (1) Customer Identity Verification &  &  &  & \checkmark \\
    \hline
    (2) High-Risk/Suspicious Transaction Monitoring  &  &  &  &  \\
    \hline
    (3) Detection of Changes to Customer Identity Information  &  &  &  & \checkmark \\
    \hline
    (4) Contract Version Tracking &  &  & \checkmark & \checkmark \\
    \hline
    (5) Exploration of Transaction History by Asset Type  &  &  &  &  \\
    \hline
    (6) External Audit   &  &  &  &  \\
    \hline
    (7) Setting Role-Based Permissions  &  &  & \checkmark & \checkmark \\
    \hline
    (8) Asset Freeze  &  & \checkmark & \checkmark & \checkmark \\
    \hline
    (9) Asset Recovery   &  & \checkmark & \checkmark & \checkmark \\
    \hline
    (10) Trading Restrictions &  & \checkmark & \checkmark & \checkmark \\
    \hline
    (11) Transaction Limit  &  &  &  &  \\
    \hline
    (12) Cancellation or Modification of Transactions  &  &  &  &  \\
    \hline
    (13) Pausing of Trading  &  &  & \checkmark & \checkmark \\
    \hline
    (14) Suspension or Disposal of Smart Contract (kill switch)  &  &  &  & \checkmark \\
    \hline
    (15) Blacklist Management  &  &  &  &  \\
    \hline
    (16) Forced Liquidation &  & \checkmark & \checkmark & \checkmark    \\
    \hline
    (17) Privacy of Personal Information &  &  &  & \checkmark    \\
    \hline
    (18) Privacy of Financial Transactions(Data) &  &  &  &     \\
    \hline
    (19) Code Security &  &  &  &     \\
    \hline
    (20) Immutability of the Ledger & \checkmark & \checkmark & \checkmark & \checkmark    \\
    \hline
    (21) Finality of Transactions and Payments & \checkmark & \checkmark & \checkmark & \checkmark    \\
    \hline
    (22) Attaching Legal Documents &  &  & \checkmark &     \\
    \hline
    (23) Token Expired Time &  &  &  &     \\
    \hline
    (24) Token Transfer Restrictions &  & \checkmark & \checkmark & \checkmark    \\
    \hline
    (25) Issuance of Tokenized Cash	  & \checkmark & \checkmark & \checkmark & \checkmark    \\
    \hline
    (26) Issuance of Tokenized Securities  &  &  & \checkmark & \checkmark    \\
    \hline
    (27) Controlling Transactions Involving Splitting Below Decimal Units &  &  &  &     \\
    \hline
    (28) Token Burning &  &  & \checkmark & \checkmark    \\
    \hline
    (29) Gasless Support &  &  &  &     \\
    \hline
    (30) Asset Class Management &  &  & \checkmark &     \\
    \hline
    (31) Token Supply Control &  &  & \checkmark & \checkmark    \\

    \bottomrule
  \end{tabular}

  \label{tab:table}
   \caption{Status of Regulatory Compliance with ERC Standard Protocols}

\end{table*}

ERC-20, as a general-purpose fungible token standard, addresses only the most basic of the Finality and Tokenizability functions in its base specification, providing a mechanism for only 3 of the 31 regulatory recommendations and control functions of RCP. ERC-7943, the deliberately minimal uRWA interface, concentrates entirely on enforcement chokepoints: layered over a fungible base token it addresses 8 of the 31 items, standardizing asset freeze, forced transfer, and transfer gating while inheriting the ledger-level guarantees of its base token. Its single forced-transfer primitive is intentionally neutral as to legal effect, so it cannot by itself record whether a given forced transfer is a recovery, a confiscation, or a liquidation, which is precisely the distinction RCP's action vocabulary supplies. ERC-1400, being built upon ERC-20, incorporates partitioned ownership, transfer restrictions, and document management into its base specification, providing a mechanism for 16 of the 31 items, roughly half of the overall recommendations by financial institutions and regulatory bodies. ERC-3643, a more recent security-token standard, embeds on-chain identity (ONCHAINID) and a modular compliance and permissioning layer in its base specification, and covers 18 of the 31 items when benchmarked against RCP, the most of any existing standard examined. Taken together, these standards, and ERC-3643 in particular, already provide base-specification mechanisms for much of the on-chain-addressable enforcement and identity requirements; the items they leave unmet are predominantly those that are inherently off-chain or cross-domain in nature, such as suspicious-transaction monitoring, external audit, transaction-data privacy, and code security, which cannot be discharged at the token-standard level alone. The load-bearing signal in this comparison is therefore not the totals but the pattern they expose: requirements left unaddressed by every standard, and a single mechanism that satisfies several legally distinct actions at once, mark exactly where a regulatory reference such as RCP is needed to give on-chain enforcement its legal legibility.

\subsection{Advantage}

Due to the varied objectives and jurisdictions of global financial regulatory bodies, and the slightly different regulations each imposes, complying with all relevant regulations is not straightforward. Through our analysis, we organized the regulations of 15 institutions into recommendations and functionalities in Table 1, and used Table 2 to map how existing standards meet those recommendations against RCP as a value-neutral reference. Additionally, we compiled how well ERC-20, ERC-7943, ERC-1400, and ERC-3643 comply with the regulations of each institution in Table 3, specifically organizing the compliance of each institution's recommendations and functionalities against the total number of such criteria across all institutions. This allows us to see, institution by institution, how far each existing standard reaches toward the regulatory and control functions that financial institutions require, and where the residual gaps lie.


\begin{table}[t]
  \centering
  {\small
  \begin{tabular}{lcccc}
    \toprule
    Institution & ERC-20 & ERC-7943 & ERC-1400 & ERC-3643 \\
    \midrule
    WB & 0/3 & 0/3 & 0/3 & 2/3 \\
    ISDA  & 1/7 & 1/7 & 2/7 & 3/7 \\
    IOSCO  & 3/15 & 5/15 & 8/15 & 9/15 \\
    IMF & 0/4 & 1/4 & 1/4 & 2/4 \\
    FSB   & 0/4 & 1/4 & 1/4 & 2/4 \\
    FATF  & 1/14 & 4/14 & 7/14 & 10/14 \\
    BIS  & 2/7 & 3/7 & 4/7 & 5/7 \\
    SFC   & 1/10 & 2/10 & 3/10 & 5/10 \\
    HKMA & 2/10 & 2/10 & 5/10 & 7/10 \\
    EU  & 1/12 & 2/12 & 5/12 & 8/12 \\
    ESMA  & 1/6 & 2/6 & 4/6 & 4/6 \\
    FCA  & 1/3 & 1/3 & 1/3 & 1/3 \\
    MAS  & 0/3 & 0/3 & 1/3 & 2/3 \\
    FINMA & 0/5 & 0/5 & 1/5 & 2/5    \\
    FINRA & 1/14 & 4/14 & 7/14 & 8/14    \\
    \hline
    Total & 14/117 & 28/117 & 50/117 & 70/117    \\

    \bottomrule
    \addlinespace
  \end{tabular}
  }
  \label{tab:table}
  \caption{Regulatory Compliance Status of ERC Protocols by Regulatory Authority}

\end{table}

ERC-20, as a general-purpose token standard, provides base-specification mechanisms for almost none of the institution-level requirements, with no institution having more than half of its recommendations and functionalities addressed, leading to the understanding that no institution would accept ERC-20 alone for the tokenization of financial assets. ERC-7943 improves on ERC-20 by standardizing enforcement chokepoints, yet because it is a deliberately minimal interface it still reaches more than half of no single institution's requirements; its contribution is concentration on freeze, forced transfer, and transfer gating rather than breadth of coverage. ERC-1400 makes further progress, providing mechanisms for more than half of the regulatory functions for institutions such as IOSCO, BIS, and ESMA, and for exactly half in the cases of FATF, HKMA, and FINRA. ERC-3643 covers the institution-level requirements more broadly than the others, reaching more than half for WB, IOSCO, FATF, BIS, HKMA, EU, ESMA, MAS, and FINRA, and exactly half for IMF, FSB, and SFC, which reflects its on-chain identity and modular compliance layer addressing a larger share of the enforcement and identity recommendations in the base specification. The requirements that remain unaddressed by these standards are predominantly those that are inherently off-chain or cross-domain, and therefore cannot be discharged at the token-standard level alone. These residual requirements can be compensated through the unique features of DAML, complementing RCP's compliance, so that a token standard benchmarked against RCP and a DLT such as Canton together cover the regulatory surface that neither reaches on its own.

\subsection{Limitation}

RCP embodies public neutrality and not the perspective of any specific entity by basing itself on comprehensive recommendations and financial product guidelines from regulatory bodies. However, being dependent on the recommendations and guidelines of global financial regulatory institutions, RCP may face limitations. The regulatory specifications of institutions can be modified and added at any time to align with the evolving capital markets, necessitating updates and enhancements to RCP accordingly. Given these limitations, regular follow-up research on RCP to monitor regulations and analyze and complement amended regulations is essential. Additionally, the current financial regulations on security tokens and virtual assets are not clear, posing another limitation. Our investigation and review were conducted using publicly available regulatory documents, reports, and GitHub source code, which might introduce ambiguity in defining regulatory functions. Nonetheless, we aim to bring these issues to the forefront, anticipating innovative advancements in capital markets through tokenization. The ambiguities regarding regulations are expected to be naturally resolved through appropriate responses as direct regulations emerge with the significant development and usage of tokenization.

\section{Conclusion}

The field of asset tokenization, which innovates capital markets, lacks research and resolution of regulatory issues that form the basis for interoperability, reuse, and standard technologies. Our RCP serves as the underlying protocol for executable protocols in tokenized capital markets, standardizing the complex regulations of various regulatory bodies related to tokenized assets into groups such as Traceability, Privacy, Enforceability, Finality and Tokenizability, providing a value-neutral benchmark for meeting these standards. Benchmarking existing standards against RCP makes explicit which regulatory requirements are addressed at the token level today and which remain inherently off-chain, and the development of RCP-based tokenization services and technologies resolves the legal uncertainties of tokenized assets, thus promoting innovation in capital markets.

\normalsize
\bibliography{references}


\clearpage


\onecolumn
\appendix
\section*{Appendix}

\begin{figure*}[ht]
    \centering
    \caption{Bond Issuance and Lifecycle Management Scenario}

    \resizebox{\textwidth}{!}{%
\begin{sequencediagram}
    \newthread{issuer}{Issuer (Asset Originator)}
    \newinst{legal}{Legal Counsel (Notary)}
    \newinst{token}{Tokenization Service (RCP Based)}
    \newinst{broker}{Broker (Listing, Payment/Settlement)}
    \newinst{kyc}{KYC}
    \newinst{investor}{Investor}
    \newinst{regulator}{Regulatory Authority (Including Audit)}

    \begin{sdblock}{Alt}{RCP}
        \begin{call}{issuer}{Legal requirements and regulatory compliance review (Notary)}{legal}{}
            \begin{call}{legal}{Instructions for bond issuance and tokenization preparation}{token}{}
                \begin{sdblock}{Alt}{ERC-3643, ERC-1400}
                    \begin{call}{token}{Request for investor identity verification and change detection}{kyc}{Identity verification results}
                    \end{call}
                \end{sdblock}
            \end{call}
        \end{call}
        \begin{call}{token}{Ready and request for issuance approval}{issuer}{}
        \end{call}
    \end{sdblock}

    \begin{sdblock}{Alt}{RCP}
        \begin{call}{issuer}{Tokenization preparation instruction}{token}{}
            \begin{call}{token}{Define DAML smart contracts and deploy on DLT}{token}{}
                \begin{sdblock}{Alt}{ERC-3643, ERC-1400}
                    \begin{call}{token}{Issue Tokenized Cash (FT) and Tokenized Securities (NFT)}{token}{}
                    \end{call}
                    \begin{call}{token}{Set regulatory compliance and trading restrictions}{token}{}
                    \end{call}
                    \begin{call}{regulator}{Request asset freeze and recovery (conditional)}{token}{}
                    \end{call}
                \end{sdblock}
            \end{call}
            \begin{call}{token}{Distribute tokens to investors}{investor}{}
            \end{call}
            \begin{call}{token}{Support gasless transactions (ERC-2771)}{token}{}
            \end{call}
        \end{call}
    \end{sdblock}

    \begin{sdblock}{Alt}{RCP, ERC-3643, ERC-1400}
        \begin{call}{token}{Request advanced KYC procedure execution}{kyc}{KYC results and investor profile update}
        \end{call}
        \begin{call}{token}{Set and manage trading restrictions}{token}{}
        \end{call}
        \begin{call}{token}{Report trading restrictions and KYC status to regulatory authority}{regulator}{}
        \end{call}
        \begin{call}{regulator}{Request additional trading restrictions or guidelines (conditional)}{token}{}
        \end{call}
    \end{sdblock}

   \end{sequencediagram}
    }
\end{figure*}

\clearpage

\begin{figure*}[ht]
    \centering
    \resizebox{\textwidth}{!}{%
\begin{sequencediagram}
    \newthread{issuer}{Issuer (Asset Originator)}
    \newinst{legal}{Legal Counsel (Notary)}
    \newinst{token}{Tokenization Service (RCP Based)}
    \newinst{broker}{Broker (Listing, Payment/Settlement)}
    \newinst{kyc}{KYC}
    \newinst{investor}{Investor}
    \newinst{regulator}{Regulatory Authority (Including Audit)}

    \begin{sdblock}{Alt}{RCP}
        \begin{messcall}{investor}{Trade request}{broker}{}
        \end{messcall}
        \begin{call}{broker}{Trade verification request}{token}{}
            \begin{sdblock}{Alt}{ERC-3643, ERC-1400}
                \begin{call}{token}{Regulatory compliance check for trade}{regulator}{Trade approval}
                \end{call}
            \end{sdblock}
        \end{call}
        \begin{messcall}{token}{Trade approval notification}{broker}{}
        \end{messcall}
        \begin{messcall}{broker}{Trade completion notification}{investor}{}
        \end{messcall}
    \end{sdblock}

    \begin{sdblock}{Alt}{RCP, ERC-3643, ERC-1400}
        \begin{call}{token}{Real-time transaction monitoring and reporting}{regulator}{}
        \end{call}
        \begin{call}{regulator}{Approve or reject transactions}{token}{}
        \end{call}
        \begin{call}{token}{Maintain record immutability and audit traceability}{token}{}
        \end{call}
        \begin{call}{token}{Automated regulatory reporting}{regulator}{}
        \end{call}
    \end{sdblock}

    \begin{sdblock}{Alt}{RCP}
        \begin{call}{token}{Maturity check}{token}{}
        \end{call}
        \begin{call}{token}{Settlement preparation}{token}{}
        \end{call}
        \begin{call}{token}{Principal and interest calculation}{token}{}
        \end{call}
        \begin{call}{token}{Execute gasless settlement (ERC-2771)}{token}{}
        \end{call}
        \begin{call}{token}{Asset transfer to investors}{investor}{}
        \end{call}
        \begin{call}{token}{Settlement record and audit traceability}{token}{}
        \end{call}
        \begin{call}{token}{Regulatory reporting and transparency}{regulator}{}
        \end{call}
        \begin{call}{token}{Dispute resolution support}{token}{}
        \end{call}
    \end{sdblock}

   \end{sequencediagram}
    }

\clearpage

\end{figure*}

\clearpage

\begin{figure}[ht]
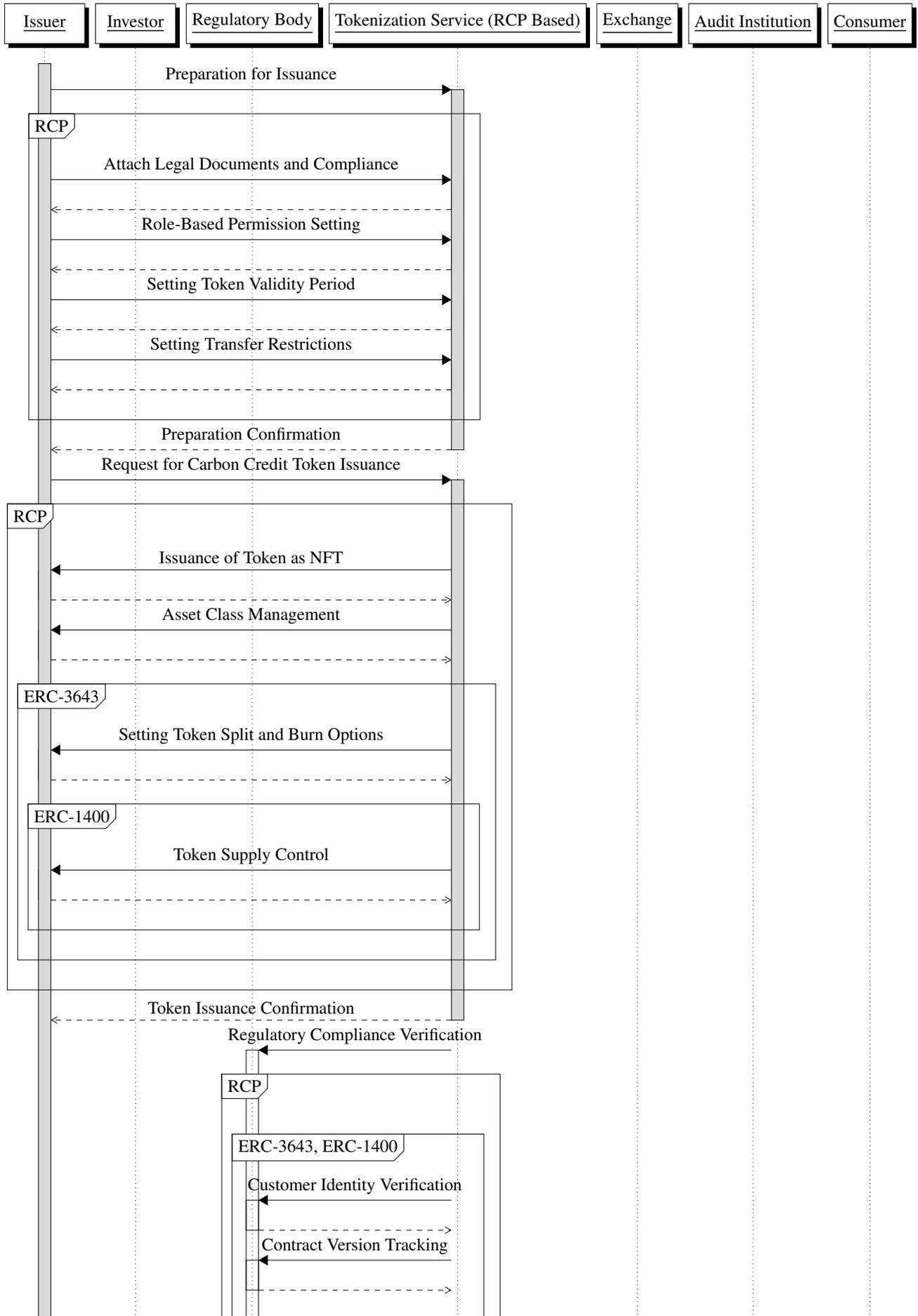

    \centering
    \caption{Carbon Credit Scenario}
    \resizebox{\textwidth}{!}{%
\begin{sequencediagram}
    \newthread{issuer}{Issuer}
    \newinst{investor}{Investor}
    \newinst{regulator}{Regulatory Body}
    \newinst{rcp}{Tokenization Service (RCP Based)}
    \newinst{exchange}{Exchange}
    \newinst{audit}{Audit Institution}
    \newinst{consumer}{Consumer}

    \begin{call}{issuer}{Preparation for Issuance}{rcp}{Preparation Confirmation}
        \begin{sdblock}{RCP}{}
            \begin{call}{issuer}{Attach Legal Documents and Compliance}{rcp}{}
            \end{call}
            \begin{call}{issuer}{Role-Based Permission Setting}{rcp}{}
            \end{call}
            \begin{call}{issuer}{Setting Token Validity Period}{rcp}{}
            \end{call}
            \begin{call}{issuer}{Setting Transfer Restrictions}{rcp}{}
            \end{call}
        \end{sdblock}
    \end{call}
    \begin{call}{issuer}{Request for Carbon Credit Token Issuance}{rcp}{Token Issuance Confirmation}
        \begin{sdblock}{RCP}{}
            \begin{call}{rcp}{Issuance of Token as NFT}{issuer}{}
            \end{call}
            \begin{call}{rcp}{Asset Class Management}{issuer}{}
            \end{call}
            \begin{sdblock}{ERC-3643}{}
                \begin{call}{rcp}{Setting Token Split and Burn Options}{issuer}{}
                \end{call}
                \begin{sdblock}{ERC-1400}{}
                    \begin{call}{rcp}{Token Supply Control}{issuer}{}
                    \end{call}
                \end{sdblock}
            \end{sdblock}
        \end{sdblock}
    \end{call}
    \begin{call}{rcp}{Regulatory Compliance Verification}{regulator}{Regulatory Compliance Confirmation}
        \begin{sdblock}{RCP}{}
            \begin{sdblock}{ERC-3643, ERC-1400}{}
                \begin{call}{rcp}{Customer Identity Verification}{regulator}{}
                \end{call}
                \begin{call}{rcp}{Contract Version Tracking}{regulator}{}
                \end{call}
            \end{sdblock}
            \begin{call}{rcp}{Blacklist Management}{regulator}{}
            \end{call}
        \end{sdblock}
    \end{call}
    \begin{call}{rcp}{Supplying Tokens to the Market}{exchange}{Market Supply Confirmation}
    \end{call}
    \begin{sdblock}{RCP}{}
        \begin{call}{consumer}{Request for Carbon Credit Use}{rcp}{Request Verification ('Using Transfer Restrictions')}
            \begin{call}{rcp}{Approval of Carbon Credit Use and Ownership Transfer}{consumer}{}
            \end{call}
            \begin{call}{rcp}{Reporting Carbon Credit Use and Attaching Legal Documents}{regulator}{Regulatory Compliance Confirmation}
            \end{call}
        \end{call}
        \begin{call}{consumer}{Request for Carbon Credit Burn}{rcp}{Burn Request Verification}
            \begin{call}{rcp}{Approval of Carbon Credit Burn and Ensuring Record Immutability}{consumer}{}
            \end{call}
            \begin{call}{rcp}{Reporting the Burn Process and Legal Compliance}{audit}{Burn Verification Result and Regulatory Compliance Reporting}
            \end{call}
        \end{call}
    \end{sdblock}
\end{sequencediagram}
}
    \label{fig:scenario}
\end{figure}

\begin{figure}[ht]
    \centering
    \resizebox{\textwidth}{!}{%
\begin{sequencediagram}
        \newthread{issuer}{Issuer}
        \newinst{investor}{Investor}
        \newinst{regulator}{Regulatory Body}
        \newinst{rcp}{Tokenization Service (RCP Based)}
        \newinst{exchange}{Exchange}
        \newinst{audit}{Audit Institution}
        \newinst{consumer}{Consumer}

    \begin{call}{investor}{Request for Purchase, Sale, or Exchange of Carbon Credit Tokens}{exchange}{Transaction Result Notification}
        \begin{call}{exchange}{Transaction Verification Request}{rcp}{Transaction Approval and Recording}
        \end{call}
    \end{call}
    \begin{sdblock}{ERC-3643, ERC-1400}{Asset Management}
        \begin{call}{issuer}{Asset Freeze, Asset Recovery Request}{rcp}{Request Approval and Recording}
        \end{call}
        \begin{call}{investor}{Token Split, Burn Request}{rcp}{Request Approval and Recording}
        \end{call}
    \end{sdblock}
    \begin{call}{regulator}{Monitoring of Transactions and Asset Management}{rcp}{Action Execution and Reporting}
    \end{call}
    \begin{sdblock}{RCP}{Market Intervention}
        \begin{call}{regulator}{Request for Temporary Suspension of Transactions, Suspension of Financial Products}{rcp}{Request Execution and Notification to the Market}
        \end{call}
        \begin{call}{regulator}{Forced Liquidation (Burn) Request}{rcp}{Request Execution and Reporting}
        \end{call}
    \end{sdblock}
    \begin{call}{audit}{Request for Audit of Transactions and Asset Management}{rcp}{Provision of Audit Information}
        \begin{call}{rcp}{Reporting Audit Results and Recommendations}{regulator}{}
        \end{call}
    \end{call}

\end{sequencediagram}
}
    \label{fig:scenario}
\end{figure}

\begin{figure}[ht]
    \centering
    \resizebox{\textwidth}{!}{%
\begin{sequencediagram}
    \newthread{issuer}{Issuer}
    \newinst{investor}{Investor}
    \newinst{regulator}{Regulatory Body}
    \newinst{rcp}{Tokenization Service (RCP Based)}
    \newinst{exchange}{Exchange}
    \newinst{audit}{Audit Institution}
    \newinst{consumer}{Consumer}

        \begin{call}{audit}{Request for Verification of Role-Based Permission Settings}{rcp}{Provision of Role-Based Permission Setting Information}
        \end{call}
        \begin{call}{audit}{Request for Verification of Legal Document Attachment and Compliance}{rcp}{Provision of Legal Document and Compliance Information}
        \end{call}
        \begin{sdblock}{Audit and Verification}{}
            \begin{call}{audit}{Request for Transaction Records and Activity Logs}{rcp}{Provision of Transaction Records and Activity Logs}
            \end{call}
            \begin{call}{audit}{Request for Verification of Record Immutability and Transaction Finality}{rcp}{Provision of Verification Results}
            \end{call}
            \begin{call}{audit}{Request for Customer Identity Verification and Transaction Restriction Verification}{rcp}{Provision of Verification Results and Related Information}
            \end{call}
            \begin{call}{audit}{Request for Verification of Asset Freeze and Blacklist Management}{rcp}{Provision of Verification Results and Related Information}
            \end{call}
            \begin{call}{audit}{Request for Verification of Asset Recovery and Forced Liquidation (Burn) Processes}{rcp}{Provision of Process Verification Results and Related Information}
            \end{call}
        \end{sdblock}
        \begin{call}{audit}{Request for Contract Version Tracking Information}{rcp}{Provision of Contract Version Tracking Information}
        \end{call}
        \begin{call}{audit}{Request for Detection of Customer Identity Information Changes and Blacklist Management Status}{rcp}{Provision of Current Status and Alert Logs}
        \end{call}
        \begin{call}{audit}{Request for Record Immutability of Audit Report}{rcp}{Provision of Report's Record Immutability and Digital Certification}
        \end{call}
        \begin{messcall}{audit}{Reporting Audit Results and Recommendations}{regulator}
        \end{messcall}
        
        \begin{call}{consumer}{Request for Use of Carbon Credit}{rcp}{Request Verification ('Using Transfer Restrictions')}
            \begin{call}{rcp}{Approval of Carbon Credit Use and Ownership Transfer}{consumer}{}
            \end{call}
            \begin{call}{rcp}{Reporting Carbon Credit Use and Attaching Legal Documents}{regulator}{Regulatory Compliance Confirmation}
            \end{call}
        \end{call}
        \begin{call}{consumer}{Request for Burn of Carbon Credit}{rcp}{Burn Request Verification}
            \begin{call}{rcp}{Approval of Carbon Credit Burn and Ensuring Record Immutability}{consumer}{}
            \end{call}
            \begin{call}{rcp}{Reporting the Burn Process and Legal Compliance}{audit}{Burn Verification Result and Regulatory Compliance Reporting}
            \end{call}
        \end{call}

\end{sequencediagram}
}
    \label{fig:scenario}
\end{figure}

\clearpage

\begin{figure*}[ht]
    \centering
    \caption{Interoperability Scenario between TradFi and DeFi}
    \resizebox{\textwidth}{!}{%
\begin{sequencediagram}
    \newthread{issuer}{Issuer (Traditional Financial Institution)}
    \newinst[1]{kyc}{KYC/AML}
    \newinst[1]{token}{Tokenization Service (DAML, RCP)}
    \newinst[1]{oracle}{Oraclizer (Oracle Service)}
    \newinst[1]{defi}{DeFi Platform}
    \newinst[1]{investor}{Investor}
    \newinst[1]{reg}{Regulatory \& Audit Agency}

    \begin{sdblock}{Bond Issuance}{}
    \begin{call}{issuer}{Check Regulatory Compliance Requirements}{reg}{}
    \end{call}
    \begin{call}{issuer}{Bond Information Modeling (Compliance with Privacy through DAML)}{token}{}
    \end{call}
    \begin{call}{token}{Apply RCP}{reg}{}
    \end{call}
    \begin{sdblock}{Loop}{Issue}
        \begin{callself}{issuer}{Issue}{}
        \end{callself}
    \end{sdblock}
    \end{sdblock}

    \begin{sdblock}{Tokenization}{}
    \begin{call}{token}{Verify Customer Identity and Regulatory Compliance}{kyc}{}
    \end{call}
    \begin{sdblock}{Loop}{Contract Modeling and Execution (Using DAML)}
        \begin{callself}{token}{Issue Token}{}
        \end{callself}
    \end{sdblock}
    \begin{callself}{token}{Asset Split and Burn}{}
    \end{callself}
    \begin{call}{token}{Interoperability and Atomic Processing}{defi}{}
    \end{call}
    \end{sdblock}

    \begin{sdblock}{Trading}{}
    \begin{call}{defi}{Request Tokenized Bonds}{oracle}{}
    \end{call}
    \begin{call}{oracle}{Transfer of Tokenized Bonds}{defi}{}
    \end{call}
    \begin{call}{defi}{Offer Trading of Tokenized Bonds}{investor}{}
    \end{call}
    \begin{call}{investor}{Request Trade}{defi}{}
    \end{call}
    \begin{call}{defi}{Verify Regulatory Compliance (KYC/AML)}{kyc}{}
    \end{call}
    \begin{call}{kyc}{Verification Result}{defi}{}
    \end{call}
    \begin{callself}{defi}{Execute Contract (Using DAML)}{}
    \end{callself}
    \begin{call}{defi}{Complete Trade and Transfer Tokens}{investor}{}
    \end{call}
    \begin{call}{defi}{Report Trade Record}{reg}{}
    \end{call}
    \end{sdblock}

    \begin{sdblock}{Settlement and Clearing}{}
    \begin{call}{defi}{Final Verification of Trade Compliance}{reg}{}
    \end{call}
    \begin{callself}{defi}{Execute Contract and Settlement (Using DAML)}{}
    \end{callself}
    \begin{call}{defi}{Asset Distribution and Clearing Processing}{investor}{}
    \end{call}
    \begin{call}{defi}{Report Trade and Clearing Record}{reg}{}
    \end{call}
    \end{sdblock}

\end{sequencediagram}
}
    \label{fig:scenario}
\end{figure*}

\clearpage

\begin{table*}[!hbp]
 \caption{Regulatory Provisions regarding Tokenization of Financial Instruments}
  \centering
  {\tiny
  \begin{tabular}{ccccp{10cm}}
    \toprule
    Property & RCP & Institution & document & article \\
    \midrule
    Traceability & (1) & World Bank & \cite{32WB2017DLT} & Legal and Regulatory Challenges

For adoption in the financial system, DLT systems will need to comply with Know-Your-Customer (KYC) and Customer Due Diligence (CDD) requirements in Anti-Money Laundering/Combating the Financing of Terrorism (AML/CFT) regulations 

\\
    & & FINMA & \cite{17FINMA2023Verordnung} & 5. Title: Special Regulations for Individuals under Article 1b of the Banking Act56 and Financial Intermediaries under Article 2 Paragraph 2 Subsections abis and dquater

Art. 44
3
If the contracting party is a legal entity or partnership, the financial intermediary must ascertain and document the authorization provisions of the contracting party regarding such entity, as well as verify the identity of the individuals acting on behalf of the legal entity or partnership in establishing the business relationship.

Art. 45
1 When establishing a business relationship with a natural person or a sole proprietor, the financial intermediary identifies the contracting party by examining an identification document provided by the contracting party.

Art. 55
1 All documents and information necessary for the identification of the contractual party must be fully submitted before executing transactions as part of the business relationship.
2 If it is not possible to identify the contractual party, the business agreement must be refused, or the business relationship must be terminated according to the following provisions.
 
 \\
    & & HKMA & \cite{20HKMA2017Ledger} & Annex B  Digital Identity Management on DLT
    
 Financial institutions are required to carry out the Know-Your-Customer (KYC) process as part of the onboarding process before they conduct business with a new client. As the number of regulatory requirements related to the KYC process and Anti-Money Laundering (AML) rules has grown, the incentive for financial institutions to find a cost-effective and user-friendly method to carry out the KYC process has increased. Digital Identity (D-ID) management has been identified as a possible means of streamlining the KYC process, enabling multiple banks to rely on the same shared, secure, and auditable source of digitized client information instead of having to collect and verify the information individually and repeatedly. 
 
 \\
    & & SFC & \cite{29SFC2023Guidelines} & IX. Dealing with Clients
    
9.5
 A Platform Operator should take all reasonable steps to establish the true and full identity of each of its clients, and, except for institutional and qualified corporate professional investors, each client’s financial situation, investment experience, and investment objectives. Where an account opening procedure other than a face-to- face approach is used, it should be one that satisfactorily ensures the identity of the client.
 
 \\
    & & BIS & \cite{6BIS2023Crypto} & Section 3  Policy measures on centrally managed cryptoasset activities

75.
 Many jurisdictions have specific requirements on AML/CFT and consumer protection. The former include mainly obligations to perform customer due diligence, transaction monitoring and suspicious transactions reporting (e.g. Europe, Japan, Singapore, the United Kingdom and the United States.) The latter requirements generally refer to the prevention of market abuse, the need to act fairly and professionally and in the best interests of the client.
 
 \\
    & & IMF-FSB & \cite{19IMF-FSB2023Synthesis} & 3. Comprehensive policy and regulatory response

3.3.2.
Jurisdictions should assess the risks of money laundering and terrorist financing associated with virtual asset activities and take appropriate steps to mitigate those risks. They should license or register virtual asset service providers and supervise the sector similarly to how they supervise other financial institutions. In addition, virtual asset service providers should be required to implement risk mitigation measures including customer due diligence, record-keeping, and reporting of suspicious transactions, as well as the implementation of targeted financial sanctions

\\
    & & IOSCO & \cite{22IOSCO2023Policy} & CHAPTER 5: RECOMMENDATIONS TO ADDRESS ABUSIVE BEHAVIOR
    
Recommendation 9 
 Market surveillance is an important tool for deterring and detecting fraudulent or manipulative activity in traditional financial markets, and market surveillance for crypto-asset markets should provide a similar level of protection. As with traditional financial markets, regulators should consider 
 
 – to the extent that existing frameworks do not already apply – the following when evaluating market surveillance tools, systems and controls that should apply to CASPs:

- Systems to identify malicious actors from a cyber, financial crime and market integrity standpoint.

- Requirements, in line with FATF recommendations for AML-CTF, including (amongst other things) Customer Due Diligence Requirements.

 \\
    & & FATF &\cite{15FATF2023Recommendations} & INTERPRETIVE NOTE TO RECOMMENDATION 10 

C. 5. 
When performing CDD measures in relation to customers that are legal persons or legal arrangements, financial institutions should be required to identify and verify the identity of the customer, and understand the nature of its business, and its ownership and control structure.
 \\
    & & & \cite{16FATF2021VIRTUAL} & PART FOUR: APPLICATION OF FATF STANDARDS TO VASPs AND OTHER OBLIGED ENTITIES THAT ENAGE IN OR PROVIDE COVERED VA ACTIVITIES
    
269. 
Like other obliged entities, in conducting CDD to fulfill their obligations under Recommendation 10, VASPs should obtain and verify the customer identification/verification information required under national law. Typically, required customer identification information includes information on the customer’s name and further identifiers such as physical address, date of birth, and a unique national identifier number (e.g., national identity number or passport number).

 \\

    & & EU & \cite{33EU2016GDPR} & (64)
    
 The controller should use all reasonable measures to verify the identity of a data subject who requests access, in particular in the context of online services and online identifiers. A controller should not retain personal data for the sole purpose of being able to react to potential requests.
 
 \\

    & (2) & FATF & \cite{15FATF2023Recommendations} & INTERPRETIVE NOTE TO RECOMMENDATION 20
    
3.  All suspicious transactions, including attempted transactions, should be reported regardless of the amount of the transaction.
 \\
    
    & & & \cite{16FATF2021VIRTUAL} & PART FOUR: APPLICATION OF FATF STANDARDS TO VASPs AND OTHER OBLIGED ENTITIES THAT ENAGE IN OR PROVIDE COVERED VA ACTIVITIES
    
275.
 Ongoing monitoring on a risk basis means scrutinizing transactions to determine whether those transactions are consistent with the VASP’s (or other obliged entity’s) information about the customer and the nature and purpose of the business relationship, wherever appropriate. Monitoring transactions also involves identifying changes to the customer profile (e.g., the customer’s behavior, use of products, and the amounts involved) and keeping it up-to-date, which may require the application of enhanced CDD measures. Monitoring transactions is an essential component in identifying transactions that are potentially suspicious, including in the context of VA transactions. Transactions that do not fit the behavior expected from a customer profile, or that deviate from the usual pattern of transactions, may be potentially suspicious.

276.
 Monitoring should be carried out on a continuous basis and may also be triggered by specific transactions. Where large volumes of transactions occur on a regular basis, automated systems may be the only realistic method of monitoring transactions, and flagged transactions should go through human/expert analysis to determine if such transactions are suspicious.

300. 
 Recommendation 20. VASPs and other obliged entities that engage in or provide VA activities, products, and services should have the ability to flag for further analysis any unusual or suspicious movements of funds or transactions—including those involving or relating to VAs—or activity that is otherwise indicative of potential involvement in illicit activity regardless of whether the transactions or activities are fiat-to-fiat, virtual-to-virtual, fiat-to-virtual, or virtual-to-fiat in nature.
 \\
    
    \bottomrule
  \end{tabular}
  }
  \label{tab:table}
\end{table*}

\begin{table*}[!hbp]
  \centering
  {\tiny
  \begin{tabular}{ccccp{10cm}}
    \toprule
    Property & RCP & Institution & document & article \\
    \midrule

    & & FINRA & \cite{18FINRA2018Rules} & 11800. CLOSE-OUT PROCEDURES

11891. .04 Such suspicious trading activities relate to allegations of fraud and therefore are not within the scope of the Rule 11890 Series. In this regard, members should routinely review the adequacy of their internal controls and ensure that appropriate system safeguards are in place to minimize or eliminate the potential for account intrusion.
    
\\
    & & MAS & \cite{28MAS2021Technology} & 14 Online Financial Services

14.3.1 
 The FI should implement real-time fraud monitoring systems to identify and block suspicious or fraudulent online transactions

14.3.2
 A process should be established to investigate suspicious transactions or payments and to ensure issues are adequately and promptly addressed.

14.3.3
 The FI should notify customers of suspicious activities or funds transfers above a threshold that is defined by the FI or customers. The notification should contain meaningful information such as type of transaction and payment amount, as well as instructions to report suspicious activities or unauthorized transactions.

\\
    & & EU & \cite{12EU2022(EU)2022/858} &  (50)

Operators of DLT market infrastructures should inform competent authorities of any material changes to their business plans or to their critical staff, of any evidence of cyber-attacks or other cyber-threats, fraud or serious malpractice, of any change in the information provided at the time of the initial application for specific permission, of any technical or operational difficulties, in particular those linked to the use of distributed ledger technology, and of any risks to investor protection, market integrity or financial stability that were not envisaged at the time when the specific permission was granted.
    
\\
    & & HKMA & \cite{20HKMA2017Ledger} & Annex E  Distributed Ledger Technology Security

Consider advanced analytics approaches to monitor for participants’ anomalous behaviour.
 
\\
    & & SFC & \cite{29SFC2023Guidelines} & VIII. Prevention of Market Manipulative and Abusive Activities

8.1 
 A Platform Operator should establish and implement written policies and controls for the proper surveillance of trading activities on its trading platform in order to identify, prevent and report any market manipulative or abusive trading activities.
(a) 
Identifying and detecting anomalies, which includes performing periodic independent reviews of suspicious price spikes;
(b)
Monitoring and preventing any potential use of abusive trading strategies;
(c) 
Taking immediate steps to restrict or suspend trading upon discovery of manipulative or abusive activities (for example, temporarily suspending accounts).
 
\\
    & & BIS & \cite{6BIS2023Crypto} & Section 3  Policy measures on centrally managed cryptoasset activities

75.
 Many jurisdictions have specific requirements on AML/CFT and consumer protection. The former include mainly obligations to perform customer due diligence, transaction monitoring and suspicious transactions reporting (e.g. Europe, Japan, Singapore, the United Kingdom and the United States.) The latter requirements generally refer to the prevention of market abuse, the need to act fairly and professionally and in the best interests of the client.
    
\\
    & & IMF-FSB & \cite{19IMF-FSB2023Synthesis} & 3. Comprehensive policy and regulatory response

3.3.2.
Jurisdictions should assess the risks of money laundering and terrorist financing associated with virtual asset activities and take appropriate steps to mitigate those risks. They should license or register virtual asset service providers and supervise the sector similarly to how they supervise other financial institutions. In addition, virtual asset service providers should be required to implement risk mitigation measures including customer due diligence, record-keeping, and reporting of suspicious transactions, as well as the implementation of targeted financial sanctions
    
\\
    & & IMF & \cite{21IMF2023ELEMENTS} &  E. Element 5 Develop and Enforce Prudential, Conduct, and Oversight

59. 
Countries need to monitor and mitigate the ML/TF risks related to decentralized finance (DeFi) projects and peer-to-peer (P2P) transactions.
    
\\
    & & IOSCO & \cite{22IOSCO2023Policy} & CHAPTER 5: RECOMMENDATIONS TO ADDRESS ABUSIVE BEHAVIOR

Recommendation 9
Market surveillance is an important tool for deterring and detecting fraudulent or manipulative activity in traditional financial markets, and market surveillance for crypto-asset markets should provide a similar level of protection. As with traditional financial markets, regulators should consider – to the extent that existing frameworks do not already apply – the following when evaluating market surveillance tools, systems and controls that should apply to CASPs:

- The timeliness of surveillance of transactions and orders to deter and detect market abuse.

- Systems for sharing information related to suspected market abuse between relevant crypto-asset markets.

- Systems to detect and report suspicious transactions and orders to the relevant body.
    
 \\
    & & ISDA & \cite{26ISDA2020Exchange} & Settlement

(ii) 
Significant changes in foreign exchange rates, as well as market, economic and political conditions, and consequently the value of the FX transaction and the extent of a counterparty’s credit exposure, may take placing during times when it is difficult for a counterparty to monitor or react. Smart contracts could, however, offer certain functionality that would allow counterparties to react automatically to these changes when certain conditions are fulfilled, or when defined events take place. This further reduces the risk for market participants in FX transactions and could reduce operational risks. Developers will also need to take into consideration how changes to relevant laws and regulations impacting FX transactions might impact DLT systems.
    
\\
    & & FINMA & \cite{17FINMA2023Verordnung} &  1. Title: General Provisions

Art 6 
1 Financial intermediaries owning branches abroad or operating a group of companies abroad must globally recognize, limit, and monitor their legal and reputational risks related to money laundering and terrorism financing. Specifically, they must ensure that:
a. the money laundering specialist department or another independent entity of the financial intermediary periodically conducts a risk analysis on a consolidated basis;
d. the group's compliance function regularly performs risk-based internal controls and spot checks on individual business relationships locally at branches and group companies.

Art. 20 
1 Financial intermediaries ensure effective monitoring of business relationships and transactions to identify increased risks.
2 Banks and securities dealers operate computer systems that assist in identifying transactions with increased risk (Article 14).

5. Title: Special Regulations for Individuals under Article 1b of the Banking Act56 and Financial Intermediaries under Article 2 Paragraph 2 Subsections abis and dquater

Art. 73
1 The financial intermediary establishes criteria for identifying transactions with elevated risks. They utilize an IT system for the detection and monitoring of transactions with heightened risks.
2 Transactions are considered high-risk if one or more transactions, appearing to be connected, reach or exceed the amount of 5000 Swiss francs.

\\
    & (3) & FATF & \cite{15FATF2023Recommendations} & INTERPRETIVE NOTE TO RECOMMENDATION 10 

A. 1.  
If, during the establishment or course of the customer relationship, or when conducting occasional transactions, a financial institution suspects that transactions relate to money laundering or terrorist financing, then the institution should: 
(a) 
normally seek to identify and verify the identity of the customer and the beneficial owner, whether permanent or occasional, and irrespective of any exemption or any designated threshold that might otherwise apply
\\
    \bottomrule
  \end{tabular}
  }
  \label{tab:table}
\end{table*}

\begin{table*}[!hbp]
  \centering
  {\tiny
  \begin{tabular}{ccccp{10cm}}
    \toprule
    Property & RCP & Institution & document & article \\
    \midrule
    & & & \cite{16FATF2021VIRTUAL} & PART THREE: APPLICATION OF FATF STANDARDS TO COUNTRIES AND COMPETENT AUTHORITIES

160.
Countries should require VASPs and other obliged entities that engage in or provide VA products and services to keep documents, data, or information collected under the CDD process up-to-date and relevant by undertaking reviews of existing records

PART FOUR: APPLICATION OF FATF STANDARDS TO VASPs AND OTHER OBLIGED ENTITIES THAT ENAGE IN OR PROVIDE COVERED VA ACTIVITIES

275. 
Ongoing monitoring on a risk basis means scrutinizing transactions to determine whether those transactions are consistent with the VASP’s (or other obliged entity’s) information about the customer and the nature and purpose of the business relationship, wherever appropriate. Monitoring transactions also involves identifying changes to the customer profile (e.g., the customer’s behavior, use of products, and the amounts involved) and keeping it up-to-date, which may require the application of enhanced CDD measures.

\\
    & & FINRA & \cite{18FINRA2018Rules} & 6800. CONSOLIDATED AUDIT TRAIL COMPLIANCE RULE

6840. (b) 
Each Industry Member shall submit to the Central Repository any updates, additions, or other changes to the Firm Designated ID, Customer Account Information, and Customer Identifying Information for each of its Customers with an Active Account on a daily basis.

\\
    & & HKMA & \cite{20HKMA2017Ledger} & Annex B Digital Identity Management on DLT

The client may at any time access the DLT network to receive the hash entries of his or her personal information and documents for the following purposes:

- To determine whether subsequent information and document updates have been verified by the accepting bank and the corresponding hashes have been stored in the DLT network

\\
    & (4) & ESMA & \cite{8ESMA2019Initial} & IV. Actors and business models

32 
Each party who participates in the validation process has an identical up-to-date copy of the chain or public ledger, which is a record of all the transactions. Each party's copy of the ledger is updated every time a new block is found.
\\
    & & & \cite{9ESMA2022Regime} & 5 Reviewing the technical standards for pre- and post-trade

149.
 ESMA considers that corrections under the transaction reporting regime are dependent on the fact that transaction reports/files are in sequence (i.e., NEWT/CANC/NEWT), therefore DLT infrastructures that do not request the reporting exemption should have systems in place to ensure that the right sequencing is respected.
    
\\
    & (5) & FCA & \cite{14FCA2023Finalised} & Chapter 2 Guidance on Cryptoasset Financial Promotions 

2.77
 Ownership of a cryptoasset can change, for example, under certain complex yield models or arrangements. In such cases firms should clearly and prominently disclose the changes to legal and beneficial ownership of the cryptoasset before a consumer proceeds to enter into a relevant agreement. In particular, firms should clearly and prominently disclose ‘who’ owns the legal and beneficial rights to the cryptoasset as part of the financial promotion.

\\
    & & SFC & \cite{29SFC2023Guidelines} & IX. Dealing with Clients 

9.33
 A Platform Operator should provide to each client timely and meaningful information about the transactions conducted with the client or on the client’s behalf, the client’s holdings and movements of client virtual assets and fiat currencies, and other activities in the client’s account. Where contract notes, statements of account and receipts are provided by a Platform Operator to a client, the Platform Operator should ensure that the information included in the contract notes, statements of account and receipts is fit for purpose, comprehensive and accurate in respect of the particular type of virtual asset involved.
    
\\
    & & IOSCO & \cite{22IOSCO2023Policy} & CHAPTER 7: RECOMMENDATIONS ON CUSTODY OF CLIENT MONIES AND ASSETS
    
Recommendation 12 
As is the case with traditional financial assets, regulators should set out expectations that the CASP maintain accurate and up-to-date records and accounts of Client Assets that readily establish the precise nature, amount, location and ownership status of Client Assets and the clients for whom the assets are held. The records should also be maintained in such a way that they may be used as an audit trail.
    
\\
    & & ISDA & \cite{27ISDA2019MASTER} & Payments and Deliveries
    
An important task in developing technology solutions will be to identify each of these potential payment streams (each of which may result from transactions related to a different asset class), and how these payment streams might be affected by the provisions of the ISDA Master Agreement. It will also be important to identify the nature and purpose of each payment correctly, and to create appropriate records, particularly to account for tax liabilities and to facilitate audit processes.

\\
    & (6) & FINRA & \cite{18FINRA2018Rules} & 11800. CLOSE-OUT PROCEDURES

11860. (b) (3) (C) 
When it begins providing such services and annually thereafter, submits an Auditor's Report to the SEC staff, which is not deemed unacceptable by the SEC staff.

11860. (b) (4)
"Auditor's Report" shall mean a written report that is prepared by competent, independent, external audit personnel in accordance with the standards of the American Institute of Certified Public Accountants and the Information Systems Audit and Control Association and that (i) verifies the certifications contained in paragraph (b)(3)(B) above
    
\\
    & & FCA & \cite{14FCA2023Finalised} & Chapter 2 Guidance on Cryptoasset Financial Promotions 

2.56
Proof of ownership of the underlying commodity or asset, such as through disclosures, independent audits or proof of deposits. Firms should make evidence of the underlying commodity or asset available to potential consumers, such as through the consumer journey, before they make an investment.    

\\
    & & ESMA & \cite{10ESMA2024Regulation} & 6 Guidelines on the supervision practices to detect and prevent the circumvention of the reverse solicitation exemption 

26.
Third-country firms may try to circumvent the authorisation requirements under Article 59 of MiCA by various means and practices. It is therefore paramount that competent authorities closely monitor the activity, if any, of third-country firms in their respective jurisdictions. Given that crypto-asset services are almost exclusively offered and promoted online, particular emphasis should be given to the online activities of third-country firms.    

\\
    & & EU & \cite{12EU2022(EU)2022/858} & (41)

The competent authority for a DLT market infrastructure should be allowed to require an audit to ensure that the overall IT and cyber arrangements of the DLT market infrastructure are fit for purpose. The costs of the audit should be borne by the operator of the DLT market infrastructure.    
\\
    & & & \cite{35EU2014MiFIR} & Article 24 Obligation to uphold integrity of markets

Without prejudice to the allocation of responsibilities for enforcing Regulation (EU) No 596/2014, competent authorities coordinated by ESMA in accordance with Article 31 of Regulation (EU) No 1095/2010 shall monitor the activities of investment firms to ensure that they act honestly, fairly and professionally and in a manner which promotes the integrity of the market.

Article 25 Obligation to maintain records

1.
 Investment firms shall keep at the disposal of the competent authority, for five years, the relevant data relating to all orders and all transactions in financial instruments which they have carried out, whether on own account or on behalf of a client. In the case of transactions carried out on behalf of clients, the records shall contain all the information and details of the identity of the client, and the information required under Directive 2005/60/EC of the European Parliament and of the Council (1). ESMA may request access to that information in accordance with the procedure and under the conditions set out in Article 35 of Regulation (EU) No 1095/2010.
\\
    \bottomrule
  \end{tabular}
  }
  \label{tab:table}
\end{table*}

\begin{table*}[!hbp]
  \centering
  {\tiny
  \begin{tabular}{ccccp{10cm}}
    \toprule
    Property & RCP & Institution & document & article \\
    \midrule

    & & & \cite{33EU2016GDPR} & Article 41 Monitoring of approved codes of conduct

1.
 Without prejudice to the tasks and powers of the competent supervisory authority under Articles 57 and 58, the monitoring of compliance with a code of conduct pursuant to Article 40 may be carried out by a body which has an appropriate level of expertise in relation to the subject-matter of the code and is accredited for that purpose by the competent supervisory authority.

 Article 47 Binding corporate rules

1. (j)
the mechanisms within the group of undertakings, or group of enterprises engaged in a joint economic activity for ensuring the verification of compliance with the binding corporate rules. Such mechanisms shall include data protection audits and methods for ensuring corrective actions to protect the rights of the data subject. Results of such verification should be communicated to the person or entity referred to in point (h) and to the board of the controlling undertaking of a group of undertakings, or of the group of enterprises engaged in a joint economic activity, and should be available upon request to the competent supervisory authority

\\
    & & SFC & \cite{29SFC2023Guidelines} & XI. Management, Supervision and Internal Control

11.19
 A Platform Operator should establish and maintain an independent audit function to objectively examine, evaluate and report on the adequacy, effectiveness and efficiency of the Platform Operator’s and its Associated Entity’s management, operations and internal controls.

\\
    & & IMF-FSB & \cite{19IMF-FSB2023Synthesis} & 2. Implications of crypto-assets
    
2.3.2.
Criminals will continue to target perceived weaknesses in AML/CFT frameworks, especially as further new illicit financing typologies continue to emerge. Without the implementation of regulation and enforcement in line with the FATF Standards (notably through effective regulation of VASPs or enforcement of a prohibition on all or certain VA activities), criminals will continue to exploit gaps created by inconsistent or weak AML/CFT frameworks, and crypto-assets could become an even greater threat to the integrity of the global financial system. To limit these opportunities, crypto-asset service providers should be licensed or registered and comply with all applicable requirements. Even when the standards are effectively implemented, regulators will need to actively monitor market developments and emerging vulnerabilities, as well as assess illicit finance risks.

\\
     & & IMF & \cite{21IMF2023ELEMENTS} & E. Element 5: Develop and Enforce Prudential, Conduct, and Oversight Requirements to All Actors

50.
 Entities should be transparent about the activities they are carrying out, as well as key operational functions that might impact markets and consumers. In many cases, third party audits can ensure that disclosure is accurate. Regulations should grant the power to establish the scope of external audits and the standards to be followed in performing such audits.

59.
 Countries need to monitor and mitigate the ML/TF risks related to decentralized finance (DeFi) projects and peer-to-peer (P2P) transactions.

\\
     & & IOSCO & \cite{23IOSCO2003Objectives} & Part III - Issuers, Market Intermediaries, and Secondary Markets 

10.6. Accounting and Auditing Standards 
 Accounting standards should ensure that fundamental information is available. There should be comprehensive and well-defined accounting principles that are of high and internationally acceptable quality, and provide accurate and relevant information on financial performance. Regulation should be intended to ensure:
 
- An independent verification of financial statements and compliance with accounting principles through professional external auditing.

- Any audit is conducted pursuant to well-defined and internationally acceptable standards.
     
\\
     & & FINMA & \cite{17FINMA2023Verordnung} & 1. Title: General Provisions 

Art. 6 a. 
The internal monitoring bodies, especially the compliance function and internal audit, as well as, if necessary, the group's audit firm, shall have access to all information on individual business relationships at all branches and group companies

\\
    \hline
\\
   Enforceability & (7) & MAS & \cite{28MAS2021Technology} &  9 Access Control

9.1.1
The principles of ‘never alone’, 'segregation of duties', and 'least privilege' should be applied when granting staff access to information assets so that no one person has access to perform sensitive system functions. Access rights and system privileges should be granted according to the roles and responsibilities of the staff, contractors, and service providers.

\\
    & & HKMA & \cite{20HKMA2017Ledger} & Annex C BOCHK Mortgage Loan Application

3.4
The system needs to allow for distinct levels of permission. It must allow users to specify the level of confidentiality for each transaction and to correspondingly conceal identities, transaction patterns, and terms of the contract from unauthorized participants when necessary. On the other hand, partial visibility is required to allow relevant parties to perform the transaction. Besides, proper governance guidelines (e.g., regarding authorization for access to and management of documents for data privacy and auditing) need to be established for off-chain information (i.e., the full property valuation report). The hash value of the document and the key management design grant certain privileges for such access.

\\
    & & FATF & \cite{15FATF2023Recommendations} & D. PREVENTIVE MEASURES

11. 
Financial institutions should be required by law to maintain records on transactions and information obtained through the CDD measures. The CDD information and the transaction records should be available to domestic competent authorities upon appropriate authority.

\\
    & & FINMA & \cite{17FINMA2023Verordnung} & 1. Title: General Provisions
    
Art. 6 2 a. 
The internal monitoring body, in particular the compliance function and the internal audit, and if necessary, the group's audit firm, have access to all information on individual business relationships in all branches and group companies

\\
    & & EU & \cite{33EU2016GDPR} & (63)

Where possible, the controller should be able to provide remote access to a secure system which would provide the data subject with direct access to his or her personal data.
    
\\
    & (8) & FATF & \cite{15FATF2023Recommendations} & D. PREVENTIVE MEASURES

16. 
Countries should ensure that, in the context of processing wire transfers, financial institutions take freezing action and should prohibit conducting transactions with designated persons and entities, as per the obligations set out in the relevant United Nations Security Council resolutions, such as resolution 1267 (1999) and its successor resolutions, and resolution 1373(2001), relating to the prevention and suppression of terrorism and terrorist financing.

INTERPRETIVE NOTE TO RECOMMENDATION 4

C. 4.
In response to relevant information, countries should enable the FIU or other competent authority to take immediate action, directly or indirectly, to withhold consent to or suspend a transaction suspected of being related to money laundering, predicate offences, or terrorist financing. The maximum duration of this measure should be specified and allow sufficient time to analyse the transaction and for competent authorities to initiate, where appropriate, an action to freeze or seize.

INTERPRETIVE NOTE TO RECOMMENDATION 6 

A. 1.
 Recommendation 6 requires each country to implement targeted financial sanctions to comply with the United Nations Security Council resolutions that require countries to freeze, without delay, the funds or other assets, and to ensure that no funds and other assets are made available to or for the benefit of:
\\

    \bottomrule
  \end{tabular}
  }
  \label{tab:table}
\end{table*}

\begin{table*}[!hbp]
  \centering
  {\tiny
  \begin{tabular}{ccccp{10cm}}
    \toprule
    Property & RCP & Institution & document & article \\
    \midrule

    & & & \cite{16FATF2021VIRTUAL} & PART THREE: APPLICATION OF FATF STANDARDS TO COUNTRIES AND COMPETENT AUTHORITIES
    
116. 
 Recommendation 6. Countries should also freeze without delay the funds or other assets—including VAs—of designated persons or entities and ensure that no funds or other assets—including VAs—are made available to or for the benefit of designated persons or entities in relation to the targeted financial sanctions related to terrorism and TF.

\\
    & (9) & FATF & \cite{15FATF2023Recommendations} & INTERPRETIVE NOTE TO RECOMMENDATION 4 

C. 4. 
In response to relevant information, countries should enable the FIU or other competent authority to take immediate action, directly or indirectly, to withhold consent to or suspend a transaction suspected of being related to money laundering, predicate offences, or terrorist financing. The maximum duration of this measure should be specified and allow sufficient time to analyse the transaction and for competent authorities to initiate, where appropriate, an action to freeze or seize.

D. 8.
Countries need a comprehensive range of measures, including legislative measures, available to confiscate criminal property and property of corresponding value

INTERPRETIVE NOTE TO RECOMMENDATION 40 

B. 20.
Countries should take part in and actively support multilateral networks to better facilitate rapid and constructive international cooperation in asset recovery.

\\
    & & & \cite{16FATF2021VIRTUAL} & PART FIVE: COUNTRY EXAMPLES OF RISK-BASED APPROACH TO VIRTUAL ASSETS AND VIRTUAL ASSET SERVICE PROVIDERS

346.
 U.S. departments and agencies have taken strong civil and criminal enforcement actions in both administrative proceedings and federal court to combat illicit activity relating to digital assets, such as by seeking various forms of relief, including cease and desist orders, injunctions, disgorgement with prejudgment interest, civil money penalties for willful violations and criminal sentences involving forfeiture and imprisonment. U.S. regulators and supervisors engage extensively with one another, state regulators, the DOJ, and other law enforcement agencies to support investigative and prosecutorial efforts in the digital assets space.

\\
    & (10) & FINRA & \cite{18FINRA2018Rules} & 2300. SPECIAL PRODUCTS

2360. (b) (8)
FINRA may impose from time to time such restrictions on option transactions or the exercise of option contracts in one or more series of options of any class which it determines are necessary in the interest of maintaining a fair and orderly market in option contracts, or in the underlying securities covered by such option contracts, or otherwise necessary in the public interest or for the protection of investors.

\\
    & & EU & \cite{35EU2014MiFIR} & Article 17 Access to quotes

2.
In order to limit the risk of exposure to multiple transactions from the same client, systematic internalisers shall be allowed to limit in a non-discriminatory way the number of transactions from the same client which they undertake to enter at the published conditions. They may, in a non-discriminatory way and in accordance with Article 28 of Directive 2014/65//EU, limit the total number of transactions from different clients at the same time provided that this is allowable only where the number and/or volume of orders sought by clients considerably exceeds the norm.

\\
    & & SFC & \cite{29SFC2023Guidelines} & XI. Management, Supervision and Internal Control 

11.13
 A Platform Operator should put in place effective risk management and supervisory controls for the operation of its trading platform. These controls should include:
(a) (ii)
 immediately prevent the platform from accepting suspicious client orders
    
\\
    & & IMF-FSB & \cite{19IMF-FSB2023Synthesis} & 3. Comprehensive policy and regulatory response

3.3.4.
Some authorities might consider implementing targeted or time-bound broad restrictions to manage the risks from crypto-assets.

\\
    & & IOSCO & \cite{23IOSCO2003Objectives} & Part III - Issuers, Market Intermediaries, and Secondary Markets

10.4.  
Regulators also need to give careful consideration to the circumstances in which it may be necessary for the proper functioning of the market to allow something less than full disclosure: for example, of trade secrets or incomplete negotiations. In the limited circumstances where the market requires some derogation from the objective of full and timely disclosure, there may need to be temporary suspensions from trading or restrictions on the trading activities of those who possess more complete information. In such circumstances, trading should be prohibited in the absence of full disclosure.
    
\\
    & & FATF & \cite{15FATF2023Recommendations} & INTERPRETIVE NOTE TO RECOMMENDATION 19 
    
2.
 Examples of the countermeasures that could be undertaken by countries include the following, and any other measures that have a similar effect in mitigating risks:
(e)
 Limiting business relationships or financial transactions with the identified country or persons in that country.

\\
    & (11) & FATF & \cite{15FATF2023Recommendations} & INTERPRETIVE NOTE TO RECOMMENDATION 10
    
H. 22.
The designated threshold for occasional transactions under Recommendation 10 is USD/EUR 15,000.

INTERPRETIVE NOTE TO RECOMMENDATION 16

B. 5.
Countries may adopt a de minimis threshold for cross-border wire transfers (no higher than USD/EUR 1,000)

INTERPRETIVE NOTE TO RECOMMENDATION 22

2.
Casinos should implement Recommendation 10, including identifying and verifying the identity of customers, when their customers engage in financial transactions equal to or above USD/EUR 3,000.

INTERPRETIVE NOTE TO RECOMMENDATIONS 22 AND 23

1.
 The designated thresholds for transactions are as follows:

- Casinos (under Recommendation 22) - USD/EUR 3,000 

- For dealers in precious metals and dealers in precious stones when engaged in any cash transaction (under Recommendations 22 and 23) - USD/EUR 15,000. 

\\
    & & & \cite{16FATF2021VIRTUAL} & PART FOUR: APPLICATION OF FATF STANDARDS TO VASPs AND OTHER OBLIGED ENTITIES THAT ENAGE IN OR PROVIDE COVERED VA ACTIVITIES
    
266.
Recommendation 10 also describes the scenarios under which FIs must undertake CDD measures, including in the context of establishing business relations, carrying out occasional transactions above the designated threshold (USD/EUR 1 000 for VA transactions), carrying out occasional transactions that are wire transfers as set forth under Recommendation 16 and its Interpretive Note (also USD/EUR 1 000 for VA transfers), where there is a suspicion of ML/TF, or when the FI doubts the veracity or adequacy of previously obtained customer identification data.

267.
 Although the designated thresholds above which casinos and dealers in precious metals and stones must conduct CDD for occasional transactions and for occasional transactions that are wire transfers are USD/EUR 3 000 and USD/EUR 15 000 respectively, when DNFBPs engage in any covered VA or VASP activities, they are subject to the CDD standards as set forth under INR. 15 (i.e., a de minimis threshold of USD/EUR 1 000 for occasional transactions and for occasional transactions that are wire transfers).

\\
    & & SFC & \cite{29SFC2023Guidelines} & IX. Dealing with Clients
    
9.7
Except for institutional and qualified corporate professional investors, a Platform Operator should set a limit for each client to ensure that the client’s exposure to virtual assets is reasonable, with reference to the client’s financial situation (including the client’s net worth) and personal circumstances.

XI. Management, Supervision and Internal Control 

11.13 
A Platform Operator should put in place effective risk management and supervisory controls for the operation of its trading platform. These controls should include: 
(b) (i)
prevent the entry of any orders that would exceed the limits prescribed for each client, including exposure limit referred to under paragraph 9.7 above;
\\
    \bottomrule
  \end{tabular}
  }
  \label{tab:table}
\end{table*}

\begin{table*}[!hbp]
  \centering
  {\tiny
  \begin{tabular}{ccccp{10cm}}
    \toprule
    Property & RCP & Institution & document & article \\
    \midrule

    & & ISDA & \cite{26ISDA2020Exchange} & Issues for technology developers to consider

(iii) b.
Transaction valuations are also subject to disputes, which can affect collateral transfers, and this can be another reason to consider the use of an oracle (see above). The complexity of valuations and collateral transfers generally result in collateral being transferred once per day. This can create intraday credit risk between counterparties. A potential benefit of a digitized process is that collateral transfers could occur more frequently, reducing the systemic risk, and reducing the requirement for counterparties to post initial margin and independent amounts to one another.

\\
    & & FINMA & \cite{17FINMA2023Verordnung} & 1. Title: General Provisions
    
Art. 12 2 
a. Payment instruments used for the cashless payment of goods and services and for cash withdrawals, where the stored electronic balance serves as the condition of the transaction, cannot be used to settle amounts exceeding 10,000 francs per transaction partner per month or to withdraw cash in such amounts.
b. For payment instruments charged after the transaction, the limit for cashless payments of goods and services and for cash withdrawals cannot exceed 25,000 francs per transaction partner per month.
c. For payment instruments allowing cashless payments between private individuals residing in Switzerland, the limit for transfers or receipts between individuals cannot exceed 1,000 francs per month and 5,000 francs per year.
d. For payment instruments allowing cashless payments between private individuals without residence restrictions, the limit for transfers or receipts between individuals cannot exceed 500 francs per month and 3,000 francs per year
    
\\
    & (12) & FINRA & \cite{18FINRA2018Rules} & 6700. TRADE REPORTING AND COMPLIANCE ENGINE (TRACE)

6730. (d) (4) 
Members shall append the applicable trade report modifiers or indicators as specified by FINRA to all transaction reports.
(A) 
Special Price Modifier
(B) 
Weighted Average Price Modifie

\\
    & & BIS-IOSCO & \cite{7BIS-IOSCO2012Principles} & Annex D: Summary of designs of payment systems, SSSs, and CCPs

Another key feature of a payment system’s design is the set of conditions that a payment must meet in order for it to be accepted by the system and be settled. In the most straightforward case, after the payment has been validated, the only condition for settlement is whether the sender has sufficient funds available (or access to intraday credit). If the payment exceeds the amount of funds available, the payment system may reject the payment. Alternatively, the system may temporarily place the payment in a system queue. The queued payment will be released from the queue at a later stage when all relevant conditions for settlement are satisfied.
    
\\
    & & ISDA & \cite{27ISDA2019MASTER} & CLOSE OUT AND NETTING

The ISDA Master Agreement allows either party (or, in certain scenarios, both parties) to terminate transactions entered into under the ISDA Master Agreement upon the occurrence of an event of default or termination event. As part of the close-out process, all of the outstanding payment and delivery obligations of the parties with respect to terminated transactions are replaced with a single early termination amount due from one party to the other.
    
\\
    & (13) & FINRA & \cite{18FINRA2018Rules} & 11800. CLOSE-OUT PROCEDURES
    
11892. (d)
In the event of any disruption or malfunction in the operation of the electronic communications and trading facilities of a self-regulatory organization or responsible single plan processor in connection with the transmittal or receipt of a regulatory trading halt, suspension or pause, a FINRA officer, acting on his or her own motion, shall declare as null and void any transaction in a security that occurs after the primary listing market for such security declares a regulatory trading halt, suspension or pause with respect to such security and before such regulatory trading halt, suspension or pause with respect to such security has officially ended according to the primary listing market. 

\\
    & & EU & \cite{35EU2014MiFIR} & Article 9 Waivers for non-equity instruments
    
4.
The competent authority responsible for supervising one or more trading venues on which a class of bond, structured finance product, emission allowance or derivative is traded may, where the liquidity of that class of financial instrument falls below a specified threshold, temporarily suspend the obligations referred to in Article 8. The specified threshold shall be defined on the basis of objective criteria specific to the market for the financial instrument concerned.

\\
    & & SFC & \cite{29SFC2023Guidelines} & VII. Operations

7.11
A Platform Operator should conduct ongoing monitoring of each virtual asset admitted for trading and consider whether to continue to allow it for trading (for example, whether in respect of a particular segment of its clients or whether a virtual asset continues to satisfy all the token admission criteria). Regular review reports should be submitted to the token admission and review committee. Where the committee decides to suspend or withdraw a virtual asset from trading, the Platform Operator should as soon as practicable notify clients of its decision and its rationale, inform clients holding that virtual asset of the options available, and ensure that clients are fairly treated.

7.12
Given that the specific features of a virtual asset may change throughout its life cycle, a Platform Operator should have appropriate monitoring procedures in place to keep track of any changes to a virtual asset being traded by its clients through its platform that may cause the virtual asset’s legal status to change such that the virtual asset falls within or ceases to fall within the definition of “securities” under the SFO. Should a virtual asset traded by its retail clients subsequently falls within the definition of “securities” under the SFO, the Platform Operator should cease to offer that virtual asset for trading by retail clients.

VIII. Prevention of Market Manipulative and Abusive Activities

8.1
 A Platform Operator should establish and implement written policies and controls for the proper surveillance of trading activities on its trading platform in order to identify, prevent, and report any market manipulative or abusive trading activities. The policies and controls should, at a minimum, cover the following:
(c) 
 taking immediate steps to restrict or suspend trading upon discovery of manipulative or abusive activities (for example, temporarily suspending accounts).
    
\\
    & & IOSCO & \cite{22IOSCO2023Policy} & CHAPTER 5: RECOMMENDATIONS TO ADDRESS ABUSIVE BEHAVIOR

Recommendation 9 – (Market Surveillance)
 Market surveillance is an important tool for deterring and detecting fraudulent or manipulative activity in traditional financial markets, and market surveillance for crypto-asset markets should provide a similar level of protection. As with traditional financial markets, regulators should consider – to the extent that existing frameworks do not already apply – the following when evaluating market surveillance tools, systems and controls that should apply to CASPs:
 
- Controls to take prompt remedial actions upon discovery of market abuse on their platform (e.g., suspension of trading).

\\
    & & & \cite{23IOSCO2003Objectives} & Part III - Issuers, Market Intermediaries, and Secondary Markets 

10.4.  
 Regulators also need to give careful consideration to the circumstances in which it may be necessary for the proper functioning of the market to allow something less than full disclosure: for example, of trade secrets or incomplete negotiations. In the limited circumstances where the market requires some derogation from the objective of full and timely disclosure, there may need to be temporary suspensions from trading or restrictions on the trading activities of those who possess more complete information. In such circumstances, trading should be prohibited in the absence of full disclosure.

\\
    & & FATF & \cite{15FATF2023Recommendations} & INTERPRETIVE NOTE TO RECOMMENDATION 4 

C. 4. 
 In response to relevant information, countries should enable the FIU or other competent authority to take immediate action, directly or indirectly, to withhold consent to or suspend a transaction suspected of being related to money laundering, predicate offences, or terrorist financing.
\\
    \bottomrule
  \end{tabular}
  }
  \label{tab:table}
\end{table*}

\begin{table*}[!hbp]
  \centering
  {\tiny
  \begin{tabular}{ccccp{10cm}}
    \toprule
    Property & RCP & Institution & document & article \\
    \midrule

    & & ISDA & \cite{27ISDA2019MASTER} & PAYMENTS AND DELIVERIES 

If a certain event with respect to a party has occurred (such as an event of default or a termination event), then the other party is not obliged to perform any of its obligations to make payments for as long as such event continues. The obligation to make the payment is therefore ‘suspended’ for the duration of the event. In this scenario, it is important to note that, notwithstanding the existence of grace periods in respect of certain events of default, payments and deliveries are conditional upon no event of default or potential event of default having occurred and continuing. Therefore, where an event occurs that may at some point in the future constitute an event of default with respect to a party, the other party may suspend payment. It is also important to remember that the payment obligation is suspended, rather than expunged or cancelled. If the event giving rise to the suspension is cured or ceases to exist, the payment obligation will resume on the original terms. Interest may also be payable on the payment amount that was subject to the suspension.

\\
    & (14) & EU & \cite{34EU2023(EU)2023/2854} & Article 36 Essential requirements regarding smart contracts for executing data sharing agreements

1. The vendor of an application using smart contracts or, in the absence thereof, the person whose trade, business or
profession involves the deployment of smart contracts for others in the context of executing an agreement or part of it, to
make data available shall ensure that those smart contracts comply with the following essential requirements of
(a) robustness and access control, to ensure that the smart contract has been designed to offer access control mechanisms
and a very high degree of robustness to avoid functional errors and to withstand manipulation by third parties;
(b) safe termination and interruption, to ensure that a mechanism exists to terminate the continued execution of
transactions and that the smart contract includes internal functions which can reset or instruct the contract to stop or
interrupt the operation, in particular to avoid future accidental executions;
(c) data archiving and continuity, to ensure, in circumstances in which a smart contract must be terminated or deactivated,
there is a possibility to archive the transactional data, smart contract logic and code in order to keep the record of
operations performed on the data in the past (auditability);
(d) access control, to ensure that a smart contract is protected through rigorous access control mechanisms at the
governance and smart contract layers; and
(e) consistency, to ensure consistency with the terms of the data sharing agreement that the smart contract executes.

\\
    & (15) & FATF & \cite{16FATF2021VIRTUAL} & PART FOUR: APPLICATION OF FATF STANDARDS TO VASPs AND OTHER OBLIGED ENTITIES THAT ENAGE IN OR PROVIDE COVERED VA ACTIVITIES
    
273.
If a VASP uncovers VA addresses that it has decided not to establish or continue business relations with or transact with due to suspicions of ML/TF, the VASP should consider making available its list of “blacklisted wallet addresses,” subject to the laws of the VASP’s jurisdiction. A VASP should screen its customer’s and counterparty’s wallet addresses against such available blacklisted wallet addresses as part of its ongoing monitoring. A VASP should make its own risk-based assessment and determine whether additional mitigating or preventive actions are warranted if there is a positive hit.
 
\\
    & & FINRA & \cite{18FINRA2018Rules} & 11500. DELIVERY OF SECURITIES WITH RESTRICTIONS
    
11540. (b) 
 If a specific certificate tendered in settlement of a contract in foreign securities is on a black list, blocked list, or subject to similar stoppage, from which an innocent holder in due course cannot have it removed by simple request, such certificate is not a good delivery, and reclamation may be made without limit of time.    
    
\\
    & & SFC & \cite{30SFC2020Terms} & XIII. Anti-Money Laundering / Counter-Financing of Terrorism

13.1 (g) 
 to use appropriate technology and wherever appropriate third party services to
identify the following situations and apply enhanced customer due diligence
and ongoing monitoring, and other additional mitigating or preventive actions
as necessary to mitigate the ML/TF risks involved:
(ii)
 Transactions involving tainted wallet addresses such as “darknet” marketplace transactions and those involving tumblers;
    
\\
    & (16) & FINRA & \cite{18FINRA2018Rules} & 4200. MARGIN
    
4210. (g) (14) (A) 
A member is required immediately either to liquidate, or transfer to another broker-dealer eligible to carry portfolio margin accounts, all portfolio margin accounts with positions in related instruments if the member is:
(i) 
insolvent as defined in Section 101 of Title 11 of the United States Code, or is unable to meet its obligations as they mature; 
(ii) 
the subject of a proceeding pending in any court or before any agency of the United States or any State in which a receiver, trustee, or liquidator for such debtor has been appointed;
(iii) 
not in compliance with applicable requirements under the Exchange Act or rules of the SEC or any self-regulatory organization with respect to financial responsibility or hypothecation of eligible participant's securities; or
(iv) 
unable to make such computations as may be necessary to establish compliance with such financial responsibility or hypothecation rules.

\\    
    & & BIS-IOSCO & \cite{7BIS-IOSCO2012Principles} & Principle 13: Participant-default rules and procedures 

3.13.4.
 A CCP should have rules and procedures to facilitate the prompt close out or transfer of a defaulting participant’s proprietary and customer positions. Typically, the longer these positions remain open on the books of the CCP, the larger the CCP’s potential credit exposures resulting from changes in market prices or other factors will be. A CCP should have the ability to apply the proceeds of liquidation, along with other funds and assets of the defaulting participant, to meet the defaulting participant’s obligations. It is critical that a CCP has the authority to act promptly to contain its exposure, while having regard for overall market effects, such as sharp declines in market prices.

\\
    \hline
\\
    Privacy & (17) & FATF & \cite{15FATF2023Recommendations} & INTERPRETIVE NOTE TO RECOMMENDATION 40
    
A. 2. 
Countries should not prohibit or place unreasonable or unduly restrictive conditions on the provision of exchange of information or assistance.
4.
Competent authorities should maintain appropriate confidentiality for any request for cooperation and the information exchanged, in order to protect the integrity of the investigation or inquiry, consistent with both parties’ obligations concerning privacy and data protection. At a minimum, competent authorities should protect exchanged information in the same manner as they would protect similar information received from domestic sources.
    
\\
    & & FINRA & \cite{18FINRA2018Rules} & 12300. Filing and Serving Documents

12300. (d) (1) (A) 
In filings with the Director, a party must redact any document that contains an individual's Social Security number, taxpayer identification number or financial account number to include only the last four digits of any of these numbers.
    
\\
    & & HKMA & \cite{20HKMA2017Ledger} & Annex E Distributed Ledger Technology Security

 In addition to addressing the confidentiality of protected information stored in the DLT, it is important to consider the confidentiality of metadata stored in DLT. In addition to transactions being stored transparently, public keys that transact are anonymous but fixed, meaning that transactions and transaction participants can be easily tracked over time. Applying advanced analytics approaches to that data could also lead to de-identification of participants and creation of new sensitive data. To further exacerbate the problem, many jurisdictions are implementing the “right to be forgotten” laws providing consumers an option to request their personal information to be removed from the databases.    
    
\\
    & & IOSCO & \cite{22IOSCO2023Policy} & CHAPTER 5: RECOMMENDATIONS TO ADDRESS ABUSIVE BEHAVIORS

Recommendation 10 
Regulators should require a CASP to put in place systems, policies, and procedures around the management of material non-public information, including, where relevant, information related to whether a crypto-asset will be admitted or listed for trading on its platform and information related to client orders, trade execution, and personally identifying information.
\\
    \bottomrule
  \end{tabular}
  }
  \label{tab:table}
\end{table*}

\begin{table*}[!hbp]
  \centering
  {\tiny
  \begin{tabular}{ccccp{10cm}}
    \toprule
    Property & RCP & Institution & document & article \\
    \midrule

    & & ISDA & \cite{26ISDA2020Exchange} & Reporting

The requirement to protect data and information collected and to build separate silos to protect parties’ confidential information from unauthorized disclosures would also have to be considered in this context. Only information that is permitted to be disclosed to each participant in the system (e.g., CCPs, regulators, brokers, parties) should be made available to them even where data is collected centrally. Technology developers should consider designing appropriate information barriers that can be integrated into the relevant platform to address this concern.
    
\\
    & & WB & \cite{31WB2009TOWARD} & III. THE PROPOSED POLICY

B. 
17.
 While the Bank is committed to disclosing as much information as possible, there are compelling reasons to protect certain types of information. Given the Bank’s diverse roles as a development organization owned by its member countries, a financial entity, and a knowledge-based institution, the Bank needs to strike the right balance between maximum disclosure and legitimate concerns to protect “confidential” information.
(a)
 Personal information (including staff records, medical information, and personal e-mail of staff and other Bank officials), information relating to staff appointment and selection processes, information pertaining to proceedings of the Bank’s internal conflict resolution mechanisms, and information relating to investigations of allegations of staff misconduct, except to the extent permitted under the staff rules.
\\

    & & EU & \cite{33EU2016GDPR} & (26)

The principles of data protection should apply to any information concerning an identified or identifiable natural person.

Article 17 Right to erasure (‘right to be forgotten’) 

1.
 The data subject shall have the right to obtain from the controller the erasure of personal data concerning him or her without undue delay and the controller shall have the obligation to erase personal data without undue delay where one of the following grounds applies.

Article 25 Data protection by design and by default  

1.
Taking into account the state of the art, the cost of implementation and the nature, scope, context and purposes of processing as well as the risks of varying likelihood and severity for rights and freedoms of natural persons posed by the processing, the controller shall, both at the time of the determination of the means for processing and at the time of the processing itself, implement appropriate technical and organisational measures, such as pseudonymisation, which are designed to implement data-protection principles, such as data minimisation, in an effective manner and to integrate the necessary safeguards into the processing in order to meet the requirements of this Regulation and protect the rights of data subjects.

\\
    & (18) & FATF & \cite{15FATF2023Recommendations} & INTERPRETIVE NOTE TO RECOMMENDATION 40 

A. 
2. 
 Countries should not prohibit or place unreasonable or unduly restrictive conditions on the provision of exchange of information or assistance.
4.
 Competent authorities should maintain appropriate confidentiality for any request for cooperation and the information exchanged, in order to protect the integrity of the investigation or inquiry, consistent with both parties’ obligations concerning privacy and data protection. At a minimum, competent authorities should protect exchanged information in the same manner as they would protect similar information received from domestic sources.

\\
    & & HKMA & \cite{20HKMA2017Ledger} & Annex E Distributed Ledger Technology Security

In addition to addressing the confidentiality of protected information stored in the DLT, it is important to consider the confidentiality of metadata stored in DLT. In addition to transactions being stored transparently, public keys that transact are anonymous but fixed, meaning that transactions and transaction participants can be easily tracked over time. Applying advanced analytics approaches to that data could also lead to de-identification of participants and creation of new sensitive data. To further exacerbate the problem, many jurisdictions are implementing the “right to be forgotten” laws providing consumers an option to request their personal information to be removed from the databases.. 

\\
    & & ESMA & \cite{8ESMA2019Initial} & Appendix 4: Details of organizational requirements under Article 16 of MiFID 2

i)
 have sound security mechanisms to guarantee the security and authentication of the means of transfer of information, minimise risk of data corruption and unauthorized access and to prevent information leakage maintaining the confidentiality of the data at all times;
    
\\
    & & IOSCO & \cite{22IOSCO2023Policy} & CHAPTER 5: RECOMMENDATIONS TO ADDRESS ABUSIVE BEHAVIORS

Recommendation 10 
Regulators should require a CASP to put in place systems, policies, and procedures around the management of material non-public information, including, where relevant, information related to whether a crypto-asset will be admitted or listed for trading on its platform and information related to client orders, trade execution, and personally identifying information.

\\
    & & WB & \cite{31WB2009TOWARD} & Annex C. PROPOSED EXCEPTIONS

10. 
 As an organization involved in dealings on the world’s financial markets, the Bank is required to use sound financial management practices, including the maintenance of utmost prudence in the disclosure of financial information related to its activities

\\
    & (19) & EU & \cite{12EU2022(EU)2022/858} & (41)
    
DLT market infrastructures should have specific and robust IT and cyber arrangements related to the use of distributed ledger technology. Such arrangements should be proportionate to the nature, scale, and complexity of the business plan of the operator of the DLT market infrastructure. Those arrangements should also ensure the continuity and continued transparency, availability, reliability, and security of the services provided, including the reliability of any smart contracts that are used, irrespective of whether those smart contracts are created by the DLT market infrastructure itself or by a third party following outsourcing procedures. DLT market infrastructures should also ensure the integrity, security, confidentiality, availability, and accessibility of data stored on the distributed ledger. 

Article 7 Additional requirements for DLT market infrastructure

4.
 Operators of DLT market infrastructures shall ensure that the overall IT and cyber arrangements related to the use of their distributed ledger technology are proportionate to the nature, scale and complexity of their businesses. Those arrangements shall ensure the continuity and continued transparency, availability, reliability and security of their services and activities, including the reliability of smart contracts used on the DLT market infrastructure. Those arrangements shall also ensure the integrity, security and confidentiality of any data stored by those operators, and shall ensure that those data are available and accessible.
 
\\    
    \hline
\\ 
    Finality & (20) & ESMA & \cite{9ESMA2022Regime} & 4 Use of DLT for trading and settlement

30. 
 The main stated benefits put forward by respondents for the increased use of DLT-based solutions on financial markets include the following areas:

- Data integrity: data stored on the ledger has a high level of integrity, as consensus among participants is necessary to alter data blocks (subject to the distributed ledger’s specific rules).

5 Reviewing the technical standards for pre- and post-trade

147.
 The transaction reports occurred in a DLT MTF cannot be cancelled and it would not be possible to modify records in case of misreporting (more details are available in Annex 6). According to the respondents, in case of wrong report that transaction would remain available in the DLT. To correct the mistake, the entity would report a new block representing the correct transaction, but the previous one would not be eliminated.
\\
    \bottomrule
  \end{tabular}
  }
  \label{tab:table}
\end{table*}

\begin{table*}[!hbp]
  \centering
  {\tiny
  \begin{tabular}{ccccp{10cm}}
    \toprule
    Property & RCP & Institution & document & article \\
    \midrule

    & & HKMA & \cite{20HKMA2017Ledger} & Annex E Distributed Ledger Technology Security  

A mutual distributed ledger, or a blockchain, has the following key capabilities: 
Mutual - blockchains are shared across organisations, owned equally by all and dominated by no-one; 
Distributed - blockchains are inherently multi-locational data structures and any user can keep his or her own copy, thus providing resilience and robustness; 
Ledger - blockchains are immutable, once a transaction is written it cannot be erased and, along with multiple copies, this means that the ledger’s integrity can easily be proven.    
    
\\
    & & BIS-IOSCO & \cite{7BIS-IOSCO2012Principles} & Annex D: Summary of designs of payment systems, SSSs, and CCPs

A payment is final at the point in time when it becomes irrevocable and unconditional. This precise moment typically depends on the underlying legal regime and the rules of the payment system itself. In some systems, a payment becomes irrevocable as soon as the system validates it (that is, queued payment orders cannot be revoked by the sender). However, the payment may not provide funds irrevocably and unconditionally to the receiver or the beneficiary until settlement occurs and is final. In other systems, payments remain revocable until settlement takes place and, lastly, in some systems a payment can only be revoked with the receiver’s consent. In general, however, in an RTGS system, a payment becomes final after it is validated by the payment system and has passed the necessary conditionality checks.
\\

    & (21) & FINRA & \cite{18FINRA2018Rules} & 11800. CLOSE-OUT PROCEDURES

11870. (c) (1) (E)
 The carrying member and the receiving member must promptly resolve and reverse any nontransferable assets that were not properly identified during validation. In all cases, each member shall promptly update its records and bookkeeping systems and notify the customer of the action taken.

\\
    & & FCA & \cite{14FCA2023Finalised} & Chapter 2 Guidance on Cryptoasset Financial Promotions

2.77
 Ownership of a cryptoasset can change, for example, under certain complex yield models or arrangements. In such cases firms should clearly and prominently disclose the changes to legal and beneficial ownership of the cryptoasset before a consumer proceeds to enter into a relevant agreement. In particular, firms should clearly and prominently disclose ‘who’ owns the legal and beneficial rights to the cryptoasset as part of the financial promotion.

\\
    & & EU & \cite{35EU2014MiFIR} & Article 24 Obligation to uphold integrity of markets
    
    Without prejudice to the allocation of responsibilities for enforcing Regulation (EU) No 596/2014, competent authorities coordinated by ESMA in accordance with Article 31 of Regulation (EU) No 1095/2010 shall monitor the activities of investment firms to ensure that they act honestly, fairly and professionally and in a manner which promotes the integrity of the market.

\\
    & & HKMA & \cite{20HKMA2017Ledger} & Annex E Distributed Ledger Technology Security  

If any piece of information relating to any transaction is subsequently changed as a result of tampering or due to transmission errors, e.g., the exact amount of the transaction, the algorithm run on the changed block will no longer produce the correct hash and will, therefore, report an error.

\\
    & & SFC & \cite{29SFC2023Guidelines} & IX. Dealing with Clients
    
9.32 
 After a Platform Operator has effected a transaction for a client, it should confirm promptly with the client the essential features of the transaction. The following information should be included:
(a) 
 name of the virtual asset in the transaction; 
(b)
 amount or value of the transaction;
(c)
 fees and charges borne by the client including applicable exchange rates. 

\\
    & & BIS-IOSCO & \cite{7BIS-IOSCO2012Principles} & Principle 8 : Settlement finality
    
An FMI should provide clear and certain final settlement, at a minimum by the end of the value date. Where necessary or preferable, an FMI should provide final settlement intraday or in real time.

Annex D: Summary of designs of payment systems, SSSs, and CCPs 

A payment is final at the point in time when it becomes irrevocable and unconditional. This precise moment typically depends on the underlying legal regime and the rules of the payment system itself. In some systems, a payment becomes irrevocable as soon as the system validates it (that is, queued payment orders cannot be revoked by the sender). However, the payment may not provide funds irrevocably and unconditionally to the receiver or the beneficiary until settlement occurs and is final. In other systems, payments remain revocable until settlement takes place and, lastly, in some systems a payment can only be revoked with the receiver’s consent. In general, however, in an RTGS system, a payment becomes final after it is validated by the payment system and has passed the necessary conditionality checks.

\\
    & & FATF & \cite{15FATF2023Recommendations} & D. PREVENTIVE MEASURES
    
16. 
 Countries should ensure that financial institutions include required and accurate originator information, and required beneficiary information, on wire transfers and related messages, and that the information remains with the wire transfer or related message throughout the payment chain. Countries should ensure that financial institutions monitor wire transfers for the purpose of detecting those which lack required originator and/or beneficiary information, and take appropriate measures.
    
\\
    & & ISDA & \cite{27ISDA2019MASTER} & DISPUTES 

It will also be important for both technology developers and derivatives market participants to consider new types of dispute that may arise as a result of entering into smart derivatives contracts. For example, it will be important for the parties to agree upon a mechanism (whether internal or external to the smart derivatives contract) to determine or verify that any data inputs are correct, how any incorrect data inputs should be remedied, and how responsibility for errors should be apportioned. Perhaps the most fundamental issue is what happens when the commercial intent of the parties is not reflected accurately in the code. In order to avoid the risk of disputes occurring between parties, it would seem sensible to include some provision within the smart derivatives contract stating that the natural language version of the contract will prevail in the event of any inconsistencies or ensuring there is some mechanism in place to confirm, to the extent necessary, that the legal effect of any coded part of the smart derivatives contract has been appropriately validated by lawyers.

\\

    & (22) & FINRA & \cite{18FINRA2018Rules} & 5100. SECURITIES OFFERINGS, UNDERWRITING AND COMPENSATION
    
5110. (b) (4) (A) 
Unless filed by the issuer, the managing underwriter, or another member, a member that anticipates participating in a public offering of securities subject to this Rule shall file with FINRA the documents and information with respect to the offering specified in subparagraphs (5) and (6) below:

11500. DELIVERY OF SECURITIES WITH RESTRICTIONS

11540. (a) 
When the laws, regulations, rulings, instructions or orders of any government, government instrumentality or agency, or official thereof having jurisdiction, require a license, clearance certificate, affidavit of ownership or any similar document in connection with the acquisition, disposition, transfer or redemption of, or other dealing in or with respect to, any security, such security shall not be a good delivery unless accompanied by the document or documents so required.

11550. (d)
A separate (detached) assignment shall contain provision for the irrevocable appointment of an attorney, with power of substitution, and a full description of the security, including name of issuer, issue, certificate number, and amount (expressed in words and numerals).

\\    
    & & HKMA & \cite{20HKMA2017Ledger} & Annex F Innovative Application of Law to Facilitate DLT
    
Regarding digitising an original document for the DLT, the general law/practice/procedure applies regardless of it in the context of DLT or otherwise. Generally speaking, a digitised version can never receive the same legal standing as its original non-digitised version, but it is more a matter of admissibility/weight as evidence in the course of court proceedings.    
\\
    \bottomrule
  \end{tabular}
  }
  \label{tab:table}
\end{table*}

\begin{table*}[!hbp]
  \centering
  {\tiny
  \begin{tabular}{ccccp{10cm}}
    \toprule
    Property & RCP & Institution & document & article \\
    \midrule

    Tokenizability & (23) & FINRA & \cite{18FINRA2018Rules} & 2300. SPECIAL PRODUCTS

2360. (a) (14) 
The term "expiration date" of an option contract issued by The Options Clearing Corporation means the day and time fixed in accordance with the rules of The Options Clearing Corporation for the expiration of such option contracts having the same expiration month as such option contract. The term "expiration date" of all other option contracts means the date specified thereon for such.

\\
    & (24) & FINRA & \cite{18FINRA2018Rules} & 5100. SECURITIES OFFERINGS, UNDERWRITING AND COMPENSATION
    
5110. (g) (1) 
In any public equity offering, other than a public equity offering by an issuer that can meet the requirements in paragraph (b)(7)(C)(i) or (ii) any common or preferred stock, options, warrants, and other equity securities of the issuer, including debt securities convertible to or exchangeable for equity securities of the issuer, that are unregistered and acquired by an underwriter and related person during 180 days prior to the required filing date, or acquired after the required filing date of the registration statement and deemed to be underwriting compensation by FINRA, and securities excluded from underwriting compensation pursuant to paragraph (d)(5)(A), (B), (C), and (E) above, shall not be sold during the offering, or sold, transferred, assigned, pledged, or hypothecated, or be the subject of any hedging, short sale, derivative, put, or call transaction that would result in the effective economic disposition of the securities by any person for a period of 180 days immediately following the date of effectiveness or commencement of sales of the public offering, except as provided in paragraph (g)(2) below.
\\
    & & ESMA & \cite{11ESMA2024instruments} & 6 Annexes

109.
 A crypto-asset can be designed in a way that it does not allow for any transfer in capital markets. Some restrictions may be placed on negotiability by not allowing holders to negotiate and/or transfer crypto-assets to a person other than the issuer. In respect of any restrictions on the transfer of financial instruments, these need to be considered on a case-by-case basis, as the nature and impact of the restriction could be sufficient to render the instrument non-tradable

\\
    & (25) & IOSCO & \cite{24IOSCO2017Technologies} & Chapter 5: Distributed Ledger Technologies (DLT)

5.1 (v) 
Tokenization is the process of digitally representing an asset or ownership of an asset. A “token” represents an asset or ownership of an asset. Such assets can be currencies, commodities, securities or properties.
\\
    & & & \cite{25IOSCO2020Relating} & Chapter 1 - Executive Summary

In the Fintech Report, IOSCO noted that "Tokenization is the process of digitally representing an asset, or ownership of an asset. A token represents an asset or ownership of an asset. Such assets can be currencies, commodities or securities or properties.

\\
    & (26) & ESMA & \cite{11ESMA2024instruments} & 6 Annexes

134.
National competent authorities and market participants should consider that to be unique, NFTs should be considered distinct and irreplaceable where their characteristics and/or the rights they provide are not identical to the other crypto-assets issued by the same (or any other) issuer.

136.
An “interdependent value test” should be conducted by national competent authorities and market participants as part of their assessment in order to classify a crypto-asset as unique and non-fungible considering: (i) if the value of the crypto-asset primarily stems from the unique characteristics of each individual asset and the utility/benefits it offers to its holder; (ii) the extent to which the interconnection of various types of crypto-assets influences the value of one another in such a way that the NFT has no value of its own that would be decorrelated from the other NFTs in the series; as well as (iii) the unique characteristics that distinguish these crypto-assets from others.

\\

    & & FATF & \cite{16FATF2021VIRTUAL} & PART TWO: SCOPE OF FATF STANDARDS

53.
 Digital assets that are unique, rather than interchangeable, and that are in practice used as collectibles rather than as payment or investment instruments, can be referred to as non-fungible tokens (NFT) or crypto-collectibles. Such assets, depending on their characteristics, are generally not considered to be VAs under the FATF definition. However, it is important to consider the nature of the NFT and its function in practice and not what terminology or marketing terms are used. This is because the FATF Standards may cover them, regardless of the terminology. Some NFTs that on their face do not appear to constitute VAs may fall under the VA definition if they are to be used for payment or investment purposes in practice. Other NFTs are digital representations of other financial assets already covered by the FATF Standards. Such assets are therefore excluded from the FATF definition of VA, but would be covered by the FATF Standards as that type of financial asset.

\\

    & (27) & IWA & \cite{36IWA2020LINK} & A restriction on the token in that there can only be 1 whole token in the class and is not dividable.
    
\\

    & (28) & FINRA & \cite{18FINRA2018Rules} & 4200. MARGIN

4210. (g) (14) 
(A) 
A member is required immediately either to liquidate, or transfer to another broker-dealer eligible to carry portfolio margin accounts, all portfolio margin accounts with positions in related instruments if the member is:
(i) 
insolvent as defined in Section 101 of Title 11 of the United States Code, or is unable to meet its obligations as they mature;
(ii) 
the subject of a proceeding pending in any court or before any agency of the United States or any State in which a receiver, trustee, or liquidator for such debtor has been appointed;
(iii) 
not in compliance with applicable requirements under the Exchange Act or rules of the SEC or any self-regulatory organization with respect to financial responsibility or hypothecation of eligible participant's securities; or
(iv) 
unable to make such computations as may be necessary to establish compliance with such financial responsibility or hypothecation rules.
\\

    & & BIS-IOSCO & \cite{7BIS-IOSCO2012Principles} & Principle 13: Participant-default rules and procedures 

3.13.4.
 A CCP should have rules and procedures to facilitate the prompt close out or transfer of a defaulting participant’s proprietary and customer positions. Typically, the longer these positions remain open on the books of the CCP, the larger the CCP’s potential credit exposures resulting from changes in market prices or other factors will be. A CCP should have the ability to apply the proceeds of liquidation, along with other funds and assets of the defaulting participant, to meet the defaulting participant’s obligations. It is critical that a CCP has the authority to act promptly to contain its exposure, while having regard for overall market effects, such as sharp declines in market prices.

\\
    & (29) & Ronan Sandford et & \cite{4Person2020ERC-2771} & Receives signed requests off-chain from Transaction Signers and pays gas to turn it into a valid transaction that goes through a Trusted Forwarder
    
\\
    & (30) & IOSCO & \cite{22IOSCO2023Policy} & CHAPTER 7: RECOMMENDATIONS ON CUSTODY OF CLIENT MONIES AND ASSETS

Recommendation 12
 As is the case with traditional financial assets, regulators should set out expectations that the CASP maintain accurate and up-to-date records and accounts of Client Assets that readily establish the precise nature, amount, location, and ownership status of Client Assets and the clients for whom the assets are held. The records should also be maintained in such a way that they may be used as an audit trail.

\\
    & (31) & IWA & \cite{5IWA2020TAXONOMY} & Benefits of shared token standards   

The TTF is far more than just a list of definitions\ldots{}
a fungible parent token that allows new tokens to be minted, and non-fungible child token that cannot be minted, where both classes of token are transferable
    
\\

    \bottomrule
  \end{tabular}
  }
  \label{tab:table}
\end{table*}

\end{document}